\documentclass[12pt]{article}
\usepackage{amsmath}
\usepackage{amsfonts}
\usepackage{enumerate}
\usepackage{ytableau}
\usepackage{hyperref}
\usepackage{bm}
\usepackage{tikz}
\begin{document}
\def\be{\begin{equation}}
\def\bea{\begin{eqnarray}}
\def\ee{\end{equation}}
\def\eea{\end{eqnarray}}
\def\d{\partial}
\def\eps{\varepsilon}
\def\la{\lambda}
\def\b{\bigskip}
\def\nn{\nonumber \\}
\def\p{\partial}
\def\t{\tilde}
\def\h{{1\over 2}}
\def\be{\begin{equation}}
\def\bea{\begin{eqnarray}}
\def\ee{\end{equation}}
\def\eea{\end{eqnarray}}
\def\b{\bigskip}
\def\u{\uparrow}
\def \byt{\begin{ytableau}}
\def \eyt{\end{ytableau}}

\makeatletter
\def\blfootnote{\xdef\@thefnmark{}\@footnotetext}  
\makeatother

\begin{center}
{\LARGE Scalar fields on $\la$--deformed cosets}
\\
\vspace{18mm}
{\bf Oleg Lunin and Wukongjiaozi Tian}
\vspace{14mm}

Department of Physics,\\ University at Albany (SUNY),\\ Albany, NY 12222, USA\\ 

\vskip 10 mm

\blfootnote{olunin@albany.edu,~jtian2@albany.edu}

\end{center}

\begin{abstract}

We study dynamics of scalar fields on a large class of geometries described by integrable sigma models. Although equations of motion are not separable due to absence of isometries and Killing tensors, we completely determine the spectra using algebraic and group--theoretic methods.

\b

\end{abstract}

\newpage

\tableofcontents

\newpage

\section{Introduction}

Study of dynamical fields provides important insights into the structure of geometries in general relativity and string theory. Such fields satisfy Partial Differential Equations, and the main technique for solving linear PDEs involves separation of variables, which has been successfully applied to fields on black hole geometries and on space--times relevant for cosmological models. Unfortunately not all backgrounds admit separation of variables, and analyzing dynamical fields on such geometries presents an important challenge. Remarkably, in certain cases the spectrum of the scalar field can be determined by purely algebraic methods \cite{PS,PS1}\footnote{Similar ideas were also explored earlier in \cite{DVV92}.}, and the goal of  this article is to extend the result of the pioneering work \cite{PS} to more general backgrounds.

\bigskip

Separation of variables in dynamical equations is often (but not always) associated with isometries of the underlying background, and such geometric symmetries are encoded in Killing vectors. The classic example in flat space is the separation of field equations in Cartesian coordinates originating from translational invariance. Separation in spherical coordinates has the same origin, although in this case the Killing vectors do not commute. Every isometry gives rise to a conserved quantity, and one needs $(d-1)$ such charges to characterize the dynamics in $d$--dimensional space.

It turns out that equations for a scalar field might be fully separable even without a sufficient number of Killing vectors, and a well-known example of this is the Kerr black hole \cite{Kerr}. The geometry has only two $U(1)$ isometries, which are not sufficient to fully characterize dynamics in a four--dimensional spacetime. Nevertheless, equations of motion for particles and for scalar fields in the Kerr geometry are fully separable due to a third conserved quantity, which cannot be attributed to a Noether charge related to isometries \cite{Carter1}. This new integral of motion is associated with an irreducible rank-two Killing 
tensor\footnote{The Kerr metric also admits several reducible Killing tensors constructed as products of Killing vectors.} \cite{Carter1,Carter2}, which can be used to solve all dynamical equations. Over the last fifteen years these results have been extended to rotating black holes in higher dimensions \cite{Kub1,StrYano}, and it has been shown that every separation of variables and every 
set of conserved charges must come from Killing vectors or tensors. Recently Killing tensors were also used to solve the Maxwell's equations in geometries produced by rotating black holes in arbitrary dimensions \cite{LMaxw}, extending the 
classic results for the Kerr geometry \cite{Teuk}. 

Given that dynamics of the Klein--Gordon field is governed by a Partial Differential Equation, one may ask whether separation of variables is the only technique for getting analytic solutions. While the vast majority of the literature focuses on this method, a notable exception is the series of articles \cite{PS,PS1}, where the authors managed to determine the full spectrum of a scalar field on backgrounds that did not have any Killing vectors or tensors. The problem was solved using an algebraic method based on group theory, and the results clearly indicated that the wavefunctions did not separate. The goal of our article is to extend the construction of \cite{PS} to a much larger class of geometries without Killing vectors or tensors.

The approach introduced in \cite{PS} is based on mapping the problem of finding the spectrum of a scalar field into a question in Conformal Field Theory. While the background geometries discussed in \cite{PS} had no isometries, the ``hidden symmetries'' associated with CFT allowed the authors of \cite{PS} to determine the spectrum using a group theoretic construction. Specifically, article \cite{PS} studied scalar field on spaces described by gauged Wess--Zumino--Witten (WZW) models \cite{WZW} on cosets $G/H$. We briefly review this work in the beginning of section \ref{SecGroup}, and the full spectrum of the Helmholtz equation (\ref{ScalarEqn}) is given by (\ref{SfetsosEigen}). The construction of \cite{PS} is based on the group-theoretic structures underlying the coset CFT, which are not visible in the metric. A very natural question is whether this success can be extended to more general class of CFTs, which might be less special than cosets, but still admit a sufficient amount of symmetry for finding the spectrum of a scalar field. Natural candidates for such extensions are the integrable lambda--deformations of the coset CFTs introduced in \cite{SftsLmb,HMSLmb}. The goal of this article is to find the spectrum of the scalar field on such deformed backgrounds. 

Integrability is a remarkable property of field and string theories that allows one to determine the entire spectrum and to understand the dynamics using an infinite set of conserved quantities \cite{IntReview}. Although original examples of integrable string theories (or CFTs) involved highly symmetric backgrounds, such as AdS$_p\times$S$^q$ \cite{AdS5int,AdS3int,AdS2int}, over the last two decades integrability has been proven for large classes of geometries that have either very few isometries or none at all \cite{beta,eta,SftsLmb,SuperLmb,DSTLmbAdS5}. In particular, the gauged WZW models are examples of integrable string theories without geometric symmetries \cite{WZWInt}. The lambda--deformation \cite{SftsLmb,HMSLmb,SuperLmb,DSTLmbAdS5} was introduced as a one--parameter family\footnote{One can also introduce a generalized lambda deformation that depends on several parameters \cite{GenLambda}. Such families have been discussed in the literature \cite{GenLambdaMore}, but they will not be considered here.} of integrable string theories that interpolates between the gauged WZW \cite{WZW} and the Principal Chiral \cite{PCM}  models, and the resulting geometries have no isometries. One may hope that the success in applying algebraic techniques to solving the scalar equation achieved in \cite{PS} extends to this entire family, and we will demonstrate that this is indeed the case.

\bigskip

This paper has the following organization. In section \ref{SecReview} we present a brief review of so-called lambda--deformation, a one--parameter family of integrable string theories interpolating between the gauged WZW and the Principal Chiral models. In section \ref{SecUndeform} we will analyze the equation for the scalar field on the background of the $SO(6)/SO(5)$ gauged WZW model. The spectrum of eigenvalues for such field was found in article \cite{PS}, so our discussion focuses on properties of the corresponding eigenfunctions. These insights are used in section \ref{SecDeformed} to solve the eigenvalue problem on the lambda--deformed background. Section \ref{SecS3} presents the counterpart of this analysis for the 
$SO(4)/SO(3)$ coset, and section \ref{SecSN} extends the results to $SO(2n)/SO(2n-1)$. Some technical details and the explicit expressions for the wavefunctions are presented in the appendices.

\section{Brief review of the $\lambda$--deformation}
\label{SecReview}
\renewcommand{\theequation}{2.\arabic{equation}}
\setcounter{equation}{0}

In this article we are interested in finding the spectrum of the scalar field on deformed cosets, so we begin with a brief review of the lambda deformation. We refer to the original articles \cite{SftsLmb,HMSLmb} for the detailed discussion. 

\bigskip
 
Following \cite{SftsLmb}, we consider a one--parameter family of geometries which interpolate between a Principal Chiral Model (PCM) \cite{PCM} and a gauged WZW model \cite{WZW} on a coset $G/H$ while preserving integrability of strings\footnote{One can also introduce the generalized lambda deformation that preserves integrability and depends on several parameters \cite{GenLambda,GenLambdaMore}, but this article focuses on the original case introduced in \cite{SftsLmb}.}. To introduce this so--called lambda--deformation, one begins with separating the generators $T^A$ of the group $G$ into $T^a$ corresponding to the subgroup $H$ and $T^\alpha$ corresponding to the coset $G/H$, and defining the Maurer-Cartan forms on $G$ by
\bea 
&&L_\pm ^A=-i\text{Tr}(T^A g^{-1}\p_\pm g),\quad R_\pm ^A=-i\text{Tr}(T^A \p_\pm gg^{-1}),\nonumber\\
&& R^A_\mu =D_{AB}L_\mu ^B,\quad D_{AB}=\text{Tr}(T_{A}g T_{B}g^{-1}).
\eea 
Then the action of the Principal Chiral Model can be written as
\bea\label{PCPact}
S_{PCM,G/H}(g)=\frac{\kappa^2}{\pi}\int \delta_{\alpha\beta} L^{\alpha}_+ L^\beta _-.
\eea 
The $\lambda$--deformed model is constructed by performing three steps \cite{SftsLmb}:
\begin{enumerate}[(i)]
\item Consider the sum of (\ref{PCPact}) and the action of the WZW model for the same group $G$ parameterized by an element ${\hat g}$.  
\item Gauge the group $F=G$ acting by
\bea 
g\rightarrow f^{-1}g,\quad {\hat g}\rightarrow f^{-1}{\hat g}f\,.\nonumber
\eea
\item Integrate out the gauge field introduced at step (ii) and fix the gauge $g=I$.
\end{enumerate}
This procedure leads to the action
\bea\label{lmbDefrmSigma}
&&S_\lambda=S_{WZW}+\frac{k}{\pi}\int R^{T}_+M^{-1}L_-,\quad \lambda=\frac{k}{k+\kappa^2},\nonumber\\
&&M_{AB}=E_{AB}-D_{BA},~~E_{AB}=\begin{bmatrix}
	\lambda^{-1}I_{\alpha\beta}&0\\0 &I_{ab}
\end{bmatrix},
\eea 
which describes strings propagating on the geometry with the metric
\bea\label{ds2}
  ds^2&=&\frac{k}{2\pi}e_0^{\alpha}Q_{\alpha\beta} e_0^{\beta},\quad 
(e_0)_\alpha=\Big[(D_{ab}-\delta_{ab}\big)^{-1}D_{a\alpha}\Big]^T L_b-L_\alpha\,.
\eea
Here $e_0$ are the frames of the gauged WZW model, and the deformation is encoded in the matrix $Q$ defined by
\bea\label{ds2Q}
Q_{\alpha\beta}&=&\left\{I-\Big[(1-\lambda^{-1}J^{-1})^{-1}+(1-\lambda^{-1}J^{-1})^{-T}\Big]\right\}_{\alpha\beta},\\
J_{\alpha\beta}&=&D_{\alpha\beta}-D_{\alpha b}(D_{ab}-\delta_{ab})^{-1}D_{a\beta}.\nonumber
\eea 
The Kalb-Ramond field vanishes, and the dilaton is given by\footnote{Note that, in contrast to (\ref{Dil}), the dilaton describing lambda deformation of a supercoset has a nontrivial $\la$--dependence \cite{HMSLmb,SuperLmb}. In this article we focus on bosonic deformations with dilaton (\ref{Dil}) \cite{SftsLmb}.}
\bea\label{Dil}
e^{-2\Phi}=\text{Det}[D_{ab}-\delta_{ab}].
\eea 
As demonstrated in \cite{SftsLmb}, geometries (\ref{ds2}) describe a family of integrable string theories which interpolate between the gauged WZW model (for $\la=0$) and the PCM (for $\la=1$). 

In this article we are interested in excitations of the deformed geometry (\ref{ds2}), specifically, in the spectrum $\Lambda$ of the scalar field\footnote{Since the expressions for eigenvalues will appear throughout this article, we introduced a factor in the right--hand side of  (\ref{ScalarEqn}) to simplify them. In particular, as we will see later, $\Lambda$ takes integer values for the gauged WZW models.}
\bea\label{ScalarEqn}
\frac{e^{2\Phi}}{\sqrt{G}}\d_\mu\Big[e^{-2\Phi}\sqrt{G}G^{\mu\nu}\d_\nu\Psi\Big]=
-\frac{2\pi}{k}\frac{1-\la^2}{1+\la^2}\Lambda \Psi\,.\quad 
\eea
Usually such spectra are found by separation of variables which is associated either with isometries of the background or with Killing tensors\footnote{We refer to the introduction for an extensive discussion of this point.}. However, even the gauged WZW model does not have a sufficient number of isometries, so this path is not available. Interestingly, in the gWZW case, one can use the group theoretic structures to determine the spectrum completely \cite{PS}, and the goal of our article is to extend this success to the geometry (\ref{ds2}) with non--zero values of the deformation parameter $\lambda$. Before attacking this problem, it is instructive to analyze some properties of equation (\ref{ScalarEqn}) on the undeformed background, and we will do this in the next section. In section \ref{SecDeformed} the lessons from this study will be used to determine the spectrum on the deformed geometry.

\section{Scalar field on the gauged WZW model}
\label{SecUndeform}
\renewcommand{\theequation}{3.\arabic{equation}}
\setcounter{equation}{0}

Before discussing the $\la$--deformation, we need to review the spectrum of the scalar field in the standard gauged WZW model, paying particular attention to the structure of wavefunctions. As we will see in the next section, the eigenvectors of the deformed model can be constructed by taking linear combinations of the gWZW results.

\subsection{Wavefunctions from group theory}
\label{SecGroup}

Although the backgrounds corresponding to gauged WZW models have very few isometries\footnote{We will be interested in $G/H$ models, where $H$ is a maximal non--abelian subgroup. In this case the resulting gWZW geometry has no isometries \cite{PS}.}, the authors of \cite{PS} were able to determine the full spectrum of the scalar field using the group theoretical methods. In this subsection we now briefly review the construction of eigenvalues presented in \cite{PS} and study the properties of the corresponding eigenfunctions. 

\bigskip

As demonstrated in \cite{PS}, the eigenvectors of the scalar field  (\ref{ScalarEqn}) on the background described by the $G/H$ gauged WZW models can be arranged according to irreducible representations of groups $G$ and $H$ (which will be labeled as $R$ and $r$), and the spectrum of $\Lambda$ is given by \cite{PS}\footnote{A special case of this formula for the $SL(2,R)/U(1)$ coset was discovered earlier in \cite{DVV92}.}
\bea\label{SfetsosEigen}
\Lambda=k\left[\frac{C_2(R)}{k+g_G}-\frac{C_2(r)}{k+g_H}\right]\,.
\eea
Here $C_2(R)$ and $C_2(r)$ are the eigenvalues of the Casimir operators, and $(g_G,g_H)$ are the dual Coxeter numbers. Here we focus on the geometric (large $k$) limit of $G/H=SO(6)/SO(5)$, and extensions to other cosets will be discussed in sections \ref{SecS3} and \ref{SecSN}.

\bigskip

Representations of $SO(6)$ are characterized by three numbers $(L_1,L_2,L_3)$, which can be either integers or half--integers, while representations of $SO(5)$ are characterized by two such numbers. Then in the geometric limit of large $k$, the general expression (\ref{SfetsosEigen}) gives
\bea\label{UndfEigValS5}
SO(6)/SO(5):&&\Lambda=4\Big[L_1(L_1+4)+L_2(L_2+2)+L_3^2-l_1(l_1+3)-l_2(l_2+1)\Big]\,.\nn
&&L_1\ge l_1\ge L_2\ge l_2\ge |L_3|.
\eea
The corresponding wavefunctions will be denoted by
\bea\label{UndfEigFncS5}
\Psi^{[L_1,L_2,L_3]}_{[l_1,l_2]}\,.
\eea
The wavefunctions for integer and half--integer values of $(L_1,L_2,L_3;l_1,l_2)$ are constructed using slightly different methods.

\bigskip

We begin with discussing the structure of the eigenfunctions (\ref{UndfEigFncS5}) for tensor representations 
with integer values of $(L_1,L_2,L_3;l_1,l_2)$. First we recall that tensor representations of $SO(6)$ can be encoded by Young tableaux with three rows \cite{Ma}. The parameters $(L_1,L_2,|L_3|)$ of a given representation are equal to the lengths of the rows in the corresponding tableau. To incorporate information about $SO(5)$ quantum numbers $(l_1,l_2)$, we introduce modified tableaux  where index $1$ appears 
$(L_1-l_1)$ times in the first row and $(L_2-l_2)$ times in the second row \cite{Ma}. For example,
\bea\label{ExamplesMarked} 
\ytableausetup{centertableaux}
\Psi^{[100]}_{[10]}\rightarrow \byt ~ ~ \eyt\quad   \Psi^{[100]}_{[00]}\rightarrow  \byt
 1\eyt,\quad  \Psi ^{[110]}_{[10]}\rightarrow \byt
 ~\\1\eyt  \quad \Psi^{[200]}_{[10]}\rightarrow \byt
 ~&1\eyt  
\eea 
To map the modified Young tableaux into wavefunctions, we use the explicit parameterization of the $SO(6)/SO(5)$ coset introduced in \cite{Bars,DSTLmbAdS5}:
\bea\label{CosetG1}
g=\begin{bmatrix}
	1&0\\
	0&q_{ab}
\end{bmatrix} \begin{bmatrix}
	B-1&BX_b\\
	-B X_a& \delta_{ab}-B X_a X_b
\end{bmatrix}, \ B=\frac{2}{1+X^T X},\ q_{ab}=[(1+A)(1-A)^{-1}]_{ab}.
\eea
Here $X_a$ is a 5-dimensional vector, and $A$ is an antisymmetric $5\times 5$ matrix that transform under the $SO(5)$ gauge rotation $\Omega$ as
\bea\label{CosetG1a}
A\rightarrow \Omega A \Omega^{-1},\quad X\rightarrow \Omega X.
\eea 
We pick a convenient gauge where
\bea\label{CosetG3}
A=\begin{bmatrix}
     0&0&0&0&0\\
	0&0&a&0&0\\
	0&-a&0&0&0\\
	0&0&0&0&b\\
	0&0&0&-b&0
	\end{bmatrix}, \quad  {X}=(X_3,0,X_1,0,X_2).
\eea 
Then the eigenvectors (\ref{UndfEigFncS5}) become functions of five coordinates $(a,b,X_1,X_2,X_3)$, and it turns out that significant simplifications happen in new variables 
$(\alpha,\beta,x,y,Y)$ defined by
\bea\label{DefineABY}
\alpha=\frac{1}{1+a^2},\quad \beta=\frac{1}{1+b^2},\quad
x=\frac{a^2X_1^2}{1+a^2},\quad y=\frac{b^2X_2^2}{1+b^2},\quad
Y=\sqrt{1+X_1^2+X_2^2+X_3^2}.
\eea
The wavefunctions are obtained by taking a product of $L\equiv L_1+L_2+|L_3|$ matrix elements of $g$ and summing over permutations $P$ corresponding to a given Young tableau
\bea\label{TensorPrescribe}
&&\Psi^{[L_1,L_2,\pm M]}_{[l_1,l_2]}={\hat\Psi}^{[L_1,L_2,M]}_{[l_1,l_2]}\pm 
{\tilde \Psi}^{[L_1,L_2,M]}_{[l_1,l_2]}\,,\quad M=|L_3|,
\nn
\nn
&&{\hat\Psi}^{[L_1,L_2,M]}_{[l_1,l_2]}={\hat\Phi}^{[L_1,L_2,M]}_{[l_1,l_2]}-\mbox{traces}\,,\quad
{\tilde \Psi}^{[L_1,L_2,M]}_{[l_1,l_2]}={\tilde \Phi}^{[L_1,L_2,M]}_{[l_1,l_2]}-\mbox{traces}\,,\nn
&&{\hat\Phi}^{[L_1,L_2,M]}_{[l_1,l_2]}=\sum_{i_1\dots i_L}\sum_P (-1)^{\sigma(P)} 
g_{i_1 i_{P[1]}}\dots g_{i_L i_{P[L]}},\quad
L=L_1+L_2+M.\\
&&{\tilde\Phi}^{[L_1,L_2,M]}_{[l_1,l_2]}=i\sum_{i_1\dots i_L}\sum_P (-1)^{\sigma(P)} \eps^{i_1i_2i_3 j_1j_2j_3}
g_{j_1 i_{P[1]}}g_{j_2 i_{P[2]}}g_{j_3 i_{P[3]}}g_{i_4 i_{P[4]}}\dots g_{i_L i_{P[L]}},\quad
M\ge 1\nonumber
\eea
The details of this construction, as well as the explicit form of the trace contributions, are discussed in the Appendix \ref{SecAppA}. The summation over $\{i_1\dots i_L\}$ assumes that index $i_k$ can take values $(2,3,4,5,6)$ for an empty box, and $i_k$ must be equal to one for a filled box. For example,
\bea
&&{\hat\Phi}^{[100]}_{[10]}=g_{aa},\quad   \Phi^{[100]}_{[00]}=g_{11},\quad  
{\hat\Phi} ^{[110]}_{[10]}=g_{11}g_{aa}-g_{1a}g_{a1} \quad 
{\hat\Phi}^{[200]}_{[10]}=g_{11}g_{aa}+g_{1a}g_{a1}\nn
&&{\hat\Phi}^{[210]}_{[20]}=g_{aa}g_{bb}g_{11}-g_{a1}g_{bb}g_{1a}+
g_{ab}g_{ba}g_{11}-g_{a1}g_{ba}g_{1b},\\
&&{\hat\Phi}^{[111]}_{[11]}=g_{aa}g_{bb}g_{11}-g_{ab}g_{ba}g_{11}-g_{a1}g_{bb}g_{1a}-g_{aa}g_{b1}g_{1b}+g_{ab}g_{b1}g_{1a}+g_{a1}g_{ba}g_{1b}\nn
&&\tilde{\Phi}^{[111]}_{[11]}=i\epsilon^{abcdef}g_{ad}g_{be}g_{cf}.\nonumber
\eea
We imply summation over repeated indices $(a,b,\dots)$ taking values $\{2,\dots,6\}$. 
Note that  ${\hat\Psi}^{[L_1,L_2,M]}_{[l_1,l_2]}$ and ${\tilde\Psi}^{[L_1,L_2,M]}_{[l_1,l_2]}$ are degenerate eigenfunctions with the same eigenvalue (\ref{UndfEigValS5}). Furthermore due to the symmetry properties of the Young tableaux, ${\tilde\Phi}^{[L_1,L_2,0]}_{[l_1,l_2]}=0$.

Equation (\ref{TensorPrescribe}) gives an algorithmic prescription for constructing eigenfunctions corresponding to all tensor representations, but it is clear that for large quantum numbers the resulting expressions become rather complicated. Fortunately to find the {\it eigenvalues} for the deformed geometry, we need only some general properties of the wavefunctions. Here we just summarize the results, and the proofs are presented in the Appendix \ref{SecAppA}.

\bigskip
\noindent
{\bf Properties of the eigenfunction $\Psi^{[L_1,L_2,L_3]}_{[l_1,l_2]}$ for the tensor representations:}
\begin{enumerate}
\item Functions ${\hat\Psi}$ and ${\tilde\Psi}$ are polynomials of 
degree $p=L_1$ in $Y^{-2}$:
\bea\label{ConjPrprt1}
\hskip -1.2cm {\hat\Psi}&=&\frac{1}{Y^{2p}}P^{(p)}_{m_p,n_p;k}+
\frac{1}{Y^{2(p-1)}}P^{(p-1)}_{m_{p-1},n_{p-1};k}+\dots +
P^{(0)}_{m_{0},n_{0};k},\\
\label{ConjPrprt1a}
\hskip -1.2cm {\tilde\Psi}&=&\frac{\gamma}{Y^{2p}}{\tilde P}^{(p)}_{m_p,{\tilde n}_p;k}+
\frac{\gamma}{Y^{2(p-1)}}{\tilde P}^{(p-1)}_{m_{p-1},{\tilde n}_{p-1};k}+\dots +
\gamma {\tilde P}^{(0)}_{m_{0},{\tilde n}_{0};k},\quad \gamma=iabX_3
\eea 
\item Expressions $P^{(q)}_{m_q,n_q;k}$ and ${\tilde P}^{(q)}_{m_q,n_q;k}$ are polynomials in $(\alpha,\beta,x,y)$ that have degree $m_q$ in $(\alpha,\beta)$, degree $n_q$ in $(x,y)$, and degree $k$ in $(x,y,\alpha,\beta)$.
\item The power $m_q$ in the leading polynomials 
$(P^{(p)}_{m_p,n_p;k},{\tilde P}^{(p)}_{m_p,{\tilde n}_p;k})$ is $m_p=L_2+M$. 
The highest power of $\alpha$ (or $\beta$) in 
$(P^{(p)}_{m_p,n_p;k},{\tilde P}^{(p)}_{m_p,{\tilde n}_p;k})$ is $L_2$.
\label{TensorProperties} 
\item The leading polynomials $(P^{(p)}_{m_p,n_p;k},{\tilde P}^{(p)}_{m_p,{\tilde n}_p;k})$ have $n_p=l_1$, 
${\tilde n}_p=l_1-1$.  
\item The power $k$ is shared by all polynomials 
$(P^{(q)}_{m_q,n_q;k},{\tilde P}^{(q)}_{m_q,{\tilde n}_q;k})$, and it is given by $k=l_1+l_2$.
\item For large values of $(\alpha,\beta)\sim {\cal A}$ and fixed $(x,y)$, expression 
$[P^{(p)}_{m_p,n_p;k}-\gamma{\tilde P}^{(p)}_{m_p,{\tilde n}_p;k}]_{\footnotesize Y=0}$ scales as 
${\cal A}^{\bar m}$ with ${\bar m}_p=L_2-M$. The highest power of $\alpha$ (or $\beta$) 
is still  $L_2$.
\item Although the expressions $(P^{(p)}_{m_p,n_p;k},{\tilde P}^{(p)}_{m_p,n_p;k})$ with $q<p$ will not be needed for our discussion\footnote{Some explicit examples of such subleading polynomials are given in equations (\ref{tmpXX}),  (\ref{tmpXX1}), (\ref{tmpXX2}), (\ref{tmpXX3}).}, we note that as $q$ decreases, only the following options are 
possible:
\bea
m_{q-1}=\{m_q,m_{q}+1\},\quad 
n_{q-1}=\{n_q,n_{q}-1\},\quad {\tilde n}_{q-1}=\{{\tilde n}_q,{\tilde n}_{q}-1\}\,.
\eea
\end{enumerate}
For future reference we summarize the powers of various polynomials in a single equation:
\bea\label{CounPowSmry}
p=L_1,\quad k=l_1+l_2,\quad m_p=L_2+M,\quad n_p=l_1\,,\quad {\tilde n}_p=l_1-1\,.
\eea
The properties 1--7 listed above imply that in the $Y=0$ limit,
\bea\label{ScaleY}
\Psi^{[L_1,L_2,L_3]}_{[l_1,l_2]}&\sim&\frac{1}{Y^{2p}}Q(\alpha,\beta;x,y),
\eea
where at large $(\alpha,\beta)\sim {\cal A}$ and fixed $(x,y)$, function $Q$ scales as 
\bea\label{ScaleQ}
Q(\alpha,\beta;x,y)\sim {\cal A}^{L_2+L_3}.
\eea
The last relation should be contrasted with scalings in ${\hat\Psi}$ and ${\tilde\Psi}$:
\bea
P^{(p)}_{m_p,n_p;k}\sim  {\cal A}^{L_2+|L_3|},\quad 
{\tilde P}^{(p)}_{m_p,{\tilde n}_p;k}\sim  {\cal A}^{L_2+|L_3|}\,.
\eea
The limit (\ref{ScaleY})--(\ref{ScaleQ}) will be explored in the next subsection, and it will play an important 
role in extending the spectrum (\ref{UndfEigValS5}) to deformed theories in section \ref{SecDeformed}. 

\bigskip

Let us now discuss the ``spinor representations'' which are not described by the Young tableaux. The elementary building blocks are representations with 
\bea
(L_1,L_2,L_3)=(\frac{1}{2},\frac{1}{2},\pm \frac{1}{2}),\quad (l_1,l_2)=(\frac{1}{2},\frac{1}{2}),
\eea
and the corresponding wavefunctions can be determined by solving equation (\ref{ScalarEqn}) with 
$\Lambda=5$. The result reads\footnote{The linear combinations are uniquely determined by imposing the scalings (\ref{ScaleY})--(\ref{ScaleQ}) for 
$\Psi^{[\frac{1}{2},\frac{1}{2},-\frac{1}{2}]}_{[\frac{1}{2},\frac{1}{2}]}$ and requiring the symmetry $\gamma\rightarrow -\gamma$ to interchange two wavefunctions in (\ref{PsiHalf}).}
\bea\label{PsiHalf}
\Psi^{[\frac{1}{2},\frac{1}{2},\frac{1}{2}]}_{[\frac{1}{2},\frac{1}{2}]}=\frac{\sqrt{\alpha\beta}}{Y}
[1+\gamma],\quad
\Psi^{[\frac{1}{2},\frac{1}{2},-\frac{1}{2}]}_{[\frac{1}{2},\frac{1}{2}]}=\frac{\sqrt{\alpha\beta}}{Y}
[1-\gamma]
\eea
Note that both representations satisfy the counting (\ref{CounPowSmry}), (\ref{ScaleY})--(\ref{ScaleQ}). 
To extend this scaling to all representations, we recall that an arbitrary representation is characterized by five numbers
\bea\label{Lvalues}
(L_1,L_2,L_3,l_1,l_2),
\eea
which are either all integers or all half--integers, and $L_3$ can be either positive or negative. We have already shown that the wavefunctions with integer values satisfy the scaling (\ref{CounPowSmry}), (\ref{ScaleY})--(\ref{ScaleQ}). Any set (\ref{Lvalues}) with half--integer values can be decomposed in at least one of two ways:
\bea
&&(L_1,L_2,L_3,l_1,l_2)=(L'_1,L'_2,L'_3,l'_1,l'_2)+(L''_1,L''_2,L''_3,l''_1,l''_2)\nn
&&(L''_1,L''_2,L''_3,l''_1,l''_2)=
(\frac{1}{2},\frac{1}{2},\frac{1}{2},\frac{1}{2})\quad \mbox{or}\quad
(\frac{1}{2},\frac{1}{2},\frac{1}{2},-\frac{1}{2})
\nonumber
\eea
with integers $(L'_1,L'_2,L'_3,l'_1,l'_2)$. Then $\Psi^{[L_1,L_2,L_3]}_{[l_1,l_2]}$ appears as the representation with the highest weights in the product
\bea\label{HalfFusion}
 \Psi^{[L'_1,L'_2,L'_3]}_{[l'_1,l'_2]} \Psi^{[L''_1,L''_2,L''_3]}_{[l''_1,l''_2]}=\Psi^{[L_1,L_2,L_3]}_{[l_1,l_2]}+
 \dots
\eea
In particular, the highest powers of $(Y^{-2},\alpha,\beta,x,y)$ in the left--hand side must match the highest powers of $\Psi^{[L_1,L_2,L_3]}_{[l_1,l_2]}$. This shows that relations (\ref{CounPowSmry}) and (\ref{ScaleY})--(\ref{ScaleQ}) extend to representations with all allowed values of (\ref{Lvalues}). 

The procedure presented here gives an algorithmic prescription for finding 
$\Psi^{[L_1,L_2,L_3]}_{[l_1,l_2]}$, but its explicit implementation involves complicated combinatorics, and the resulting expressions are rather unwieldy. On the other hand, the eigenvalues (\ref{UndfEigValS5}) are very simple, and they can be easily extracted from looking at a particular limit of the wavefunctions, which will be discussed in the next subsection. Since in the deformed case we are mostly interested in the eigenvalues, and since the group--theoretic arguments presented here break down, we expect that the limit discussed in the next subsection would be the easiest path towards finding the spectrum on the lambda--deformed spacetimes.

\subsection{Rescaled solution}
\label{SecUndefLimit}

Let us use the properties 1-7 listed on page \pageref{TensorProperties} to define a simple limit of the wavefunctions (\ref{UndfEigFncS5}), which will eventually allow us to find the spectrum of the Helmhotlz equation (\ref{ScalarEqn}) on the deformed geometry (\ref{ds2}). 

Given the structure (\ref{ConjPrprt1}) of the wavefunction, it is convenient to isolate the highest power of $Y^{-2}$. This can be accomplished by looking at the limit of small $Y$ in the metric, while keeping the remaining coordinates $(\alpha,\beta,x,y)$ fixed. The resulting geometry turns out to be regular, although it describes an analytic continuation from the physical branch to imaginary values of $X_3$: according to (\ref{DefineABY}), small values $Y$ correspond to 
\bea
X_3\rightarrow \pm i\sqrt{1+X_1^2+X_2^2}\,.\nonumber
\eea
The limit of the wavefunction is given by (\ref{ScaleY}), then the scaling (\ref{ScaleQ}) suggests sending 
$(\alpha,\beta)$ to infinity with fixed $(x,y)$. The final sequence of limits can be summarized as
\bea\label{SeqLimits}
Y\rightarrow 0,\ \mbox{then}\ \alpha\rightarrow\infty,\ \mbox{then}\ \beta\rightarrow\infty,\ 
\mbox{fixed}\quad (x,y).
\eea
The resulting metric reads 
\bea\label{UndefGeomLimit}
ds^2=-\frac{dY^2}{Y^2}+\frac{dx d\alpha}{2x\alpha}+\frac{dy d\beta}{2y\beta}
-\frac{d\alpha^2}{4\alpha^2}\left[1+\frac{1}{x}+\frac{2y}{x}\right]
-\frac{d\beta^2}{4\beta^2}\left[1+\frac{1}{y}\right],
\eea
and it is invariant under separate rescalings of $\alpha$, $\beta$ and $Y$. 

In the limit (\ref{SeqLimits}), the solution (\ref{ConjPrprt1}) simplifies to 
\bea\label{PsiLim1}
\Psi^{[L_1,L_2,L_3]}_{[l_1,l_2]}=\frac{1}{Y^{2p}}\alpha^{L_2}\beta^{L_3}Q(x,y),
\eea
where $Q$ is some polynomial. Writing it as
\bea\label{PsiLim2}
Q(x,y)=x^\la g_0(y)+x^{\la-1}g_1(y)+\dots
\eea
and looking at the differential equation in the leading order in $x$, we find that
\bea\label{g0rescaled}
&&g_0(y)=F\Big[-\nu,1+2L_3+\nu,1,-y\Big],\\
&&\Lambda=4p(p+4)+4L_3^2+4L_2(L_2+1)-4(L_2+\la)(L_2+\la+3)-4(L_3+\nu)(L_3+\nu+1)\nonumber
\eea
For polynomial solution, both $\la$ and $\nu$ must be non--negative integers. Equations for the polynomials $(g_1(y),g_2(y),\dots)$ can be solved as well, but the resulting expressions are not very illuminating. 
The eigenvalue $\Lambda$ in (\ref{g0rescaled}) agrees with (\ref{UndfEigValS5}) upon identification
\bea\label{LmbLmapUndef}
l_1=L_2+\la,\quad l_2=L_3+\nu.
\eea
While the arguments presented in subsection \ref{SecGroup} guarantee that the eigenfunctions of the Helmholtz equation (\ref{ScalarEqn}) become polynomials in the limit (\ref{SeqLimits}), not every polynomial solution of the differential equation for $Q(x,y)$ in the metric (\ref{UndefGeomLimit}) corresponds to an eigenvector of the full problem. Some such polynomials are just artifacts of the limit, and they come from non--normalizable solutions of the equation (\ref{ScalarEqn}) on the full coset. Let us identify the values of 
$(\la,\nu)$ that lead to the relevant solutions. 

By comparing the  relations (\ref{LmbLmapUndef}) with ranges (\ref{UndfEigValS5}) for $(l_1,l_2)$, we conclude that there is a one--to--one correspondence between the positive values of $(\la,\nu)$ and the set of $(l_1,l_2)$ allowed by group theory, {\it as long as} $L_3\ge 0$. For negative values of $L_3$, relations (\ref{LmbLmapUndef}) reproduce the ranges (\ref{UndfEigValS5}) only if
\bea\label{NuIneq}
\nu\ge -2L_3.
\eea
Polynomials with positive values of $\nu$ that do not satisfy this condition are artifacts of the limiting procedure. Rewriting the solution (\ref{g0rescaled}) as
\bea\label{g0One1}
g_0(y)=(1+y)^{-2L_3}F\Big[-\nu-2L_3,1+\nu,1,-y\Big],
\eea
we observe that for $\nu$ from the range (\ref{NuIneq}), the hypergeometric function appearing in the last relation is again a polynomial, and this property distinguishes the allowed values (\ref{NuIneq}) from the artifacts of the limit. Furthermore, equation (\ref{g0One1}) can be rewritten in a very suggestive form:
\bea\label{g0One2}
g_0(y)=(1+y)^{-2L_3}F\Big[-\nu',1-2L_3+\nu,1,-y\Big],\quad \nu'=\nu+2L_3
\eea
We conclude that hypergeometric functions appearing in (\ref{g0rescaled}) and (\ref{g0One2}) are related by a simple map:
\bea\label{NuMapL3}
\nu\rightarrow\nu',\quad L_3\rightarrow -L_3\,,
\eea
which also transforms equations (\ref{LmbLmapUndef}) into their counterparts involving $\nu'$. 

It turns out that, for negative values of $L_3$, factorization into $(1+y)^{-2L_3}$ and a polynomial extends from (\ref{g0One2}) to the entire $Q(x,y)$. To see this, we observe that the metric (\ref{UndefGeomLimit}) 
is invariant under a coordinate transformation
\bea\label{BetaMapSmtry}
\beta\rightarrow \frac{(1+y)^2}{\beta}\,,
\eea
which maps (\ref{PsiLim1}) into 
\bea\label{PsiLim1map}
\Psi^{[L_1,L_2,L_3]}_{[l_1,l_2]}=
\frac{1}{Y^{2p}}\alpha^{L_2}\frac{(1+y)^{2L_3}}{\beta^{L_3}}{Q}(x,y)=
\frac{1}{Y^{2p}}\alpha^{L_2}\beta^{-L_3}\Big[(1+y)^{2L_3}{Q}(x,y)\Big]
,
\eea
with the same function $Q(x,y)$. The expression in the square brackets can be {\it interpreted} as function $Q$ for the opposite value of $L_3$. Then for $L_3<0$ we find
\bea\label{PsiLim1mapQQ}
Q^{(L_3)}(x,y)=(1+y)^{-2L_3}{Q}^{(-L_3)}(x,y).
\eea
This extend the relation (\ref{g0One2}) to the entire polynomial $Q(x,y)$. The last relation implies that to find the full spectrum of the coset model (but not all polynomial solutions of the field equations in the metric (\ref{UndefGeomLimit}) since they include some artifacts of the limit), it is sufficient to focus only on 
non--negative values of $L_3$. The remaining states can be obtained using the map (\ref{NuMapL3}), (\ref{PsiLim1map}). As we will see in the next section, restriction to non--negative $L_3$ gives the entire spectrum on the deformed geometry as well. 

The expressions (\ref{PsiLim1})--(\ref{g0rescaled}) for the limit of the wavefunction are very simple, and in the next section they will be extended to the deformed geometry, which will also allow us to determine the eigenvalues in that case.

\section{Scalar field on the deformed $SO(6)/SO(5)$ coset}
\label{SecDeformed}
\renewcommand{\theequation}{4.\arabic{equation}}
\setcounter{equation}{0}

Let us now apply the lessons we learned in the last section to the study of solutions of equation (\ref{ScalarEqn}) on the deformed background (\ref{ds2}). Specifically, we will demonstrate that even in the deformed case the eigenvectors must be polynomial functions in a particular coordinate system, and we will show that all eigenvalues can be found by solving certain {\it algebraic} equations. We begin with restricting the form of wavefunctions using group theory and the relation between the Laplace operator appearing in (\ref{ScalarEqn}) and the Hamiltonian of the sigma model (\ref{lmbDefrmSigma}). The differential equations for wavefunctions and the resulting algebraic relations for the eigenvalues will be studied in sections \ref{SecDefResc} and \ref{SecDefMatr}.

\subsection{Properties of the Laplace equation}
\label{SecDefLapl}

Our goal is to find the spectrum of the Helmholtz equation (\ref{ScalarEqn}) on the lambda--deformed geometry (\ref{ds2}). The undeformed wavefunctions discussed in the last section had a simple polynomial structure, and we will now demonstrate that the deformation mixes such solutions in a controlled way. This would imply that the deformed wavefunction is still a polynomial (although the power counting slightly differs from the one presented in the last section), so one can use the limit introduced in section \ref{SecUndefLimit} along with a simple modification of the ansatz (\ref{PsiLim1}) to determine the spectrum of the deformed theory.

\bigskip

We begin with writing the metric (\ref{ds2}) for the $SO(6)/SO(5)$ coset in the $(a,X)$ coordinates introduced by (\ref{CosetG1})--(\ref{CosetG3}). The result reads 
\bea \label{Simds}
&&ds^2=G_{\mu\nu}dx^\mu dx^\nu=
\frac{k}{2\pi }\frac{1+\lambda^2}{1-\lambda^2}(e_{(0)}e_{(0)}+\kappa e_{(0)} J e_{(0)}),
\quad \kappa=\frac{2\lambda}{1+\lambda^2},\nonumber\\
&&g_{\mu\nu }\equiv (e_{(0)}e_{(0)})_{\mu\nu },\quad h_{\mu\nu }=(e_{(0)} J e_{(0)})_{\mu\nu },\quad
 e^{-2\Phi}=\frac{ a^2 b^2 X_1^2 X_2^2 \left(a^2-b^2\right)^2}{\left(a^2+1\right)^3 \left(b^2+1\right)^3 Y^8}
\eea
Here $e_{(0)}$ are the frames of the original gauged WZW model:
\bea\label{Simds1}
e_{(0)}^1&=&\frac{X_3 da}{a(1+a^2)}+\frac{X_3 db}{b(1+b^2)}+\frac{1}{Y^2}[(Y^2-X_3^2)dX_3-X_3(X_1dX_1+X_2dX_2)],\nonumber\\ 
e_{(0)}^2&=& \frac{da}{X_1(1+a^2)},\quad e_0^4=\frac{db}{X_2 (1+b^2)},\\
e_{(0)}^3&=& \frac{(b^2-1)X_3^2-a^2(1+b^2)X_2^2}{X_1 a(1+a^2)(a^2-b^2)}da-\frac{(1+a^2)bX_1}{(1+b^2)(a^2-b^2)}db\nonumber \\ 
&&\quad +dX_1+\frac{1}{Y^2}[X_1(X_1dX_1+X_2dX_2+X_3dX_3)],\nonumber \\ 
e_{(0)}^5&=&\frac{(1+b^2)a X_2}{(1+a^2)(a^2-b^2)}da-\frac{(a^2-a)X_3-b^2(1+a^2)X_1^2}{X_2b(1+b^2)(a^2-b^2)}db\nonumber \\
&&\quad +dX_2-\frac{1}{Y^2}[X_2(X_1dX_1+X_2dX_2+X_3dX_3)].\nonumber
\eea 
The deformation parameter $\la$ appears only in the overall factor of the metric and as a part of the matrix $J$:
\be
J=\begin{bmatrix}
	1&0&0\\
	0&W(a)&0\\
	0&0&W(b)
\end{bmatrix},\quad W(x)=\begin{bmatrix}
	\frac{x^2-1}{x^2+1}&\frac{2x}{x^2+1}\\
	\frac{2x}{x^2+1}&\frac{1-x^2}{x^2+1}.
\end{bmatrix}
\ee
Using the expressions for the inverse metric and the determinant,
\bea 
G^{\mu\nu}=\frac{2\pi}{k} \frac{1+\lambda^2}{1-\lambda^2}(g^{\mu\nu }-\kappa h^{\mu\nu })\,,
\quad \sqrt{G}=\left[\frac{k}{2\pi}\right]^{5/2}\frac{(\kappa -1) (\kappa +1)^{3/2}}{\left(a^2+1\right) \left(b^2+1\right) X_1 X_2 Y^2}\,,
\nonumber
\eea 
we can decompose the Laplace operator into the undeformed part $\Delta_0$ and the deformation 
$\Delta_1$:
\bea\label{LapOpe} 
\Delta &=&\frac{1}{e^{-2\Phi}\sqrt{G}}\p_\mu(e^{-2\Phi}\sqrt{G}G^{\mu\nu }\p_\nu)=
\frac{2\pi}{k} \frac{1+\lambda^2}{1-\lambda^2}\left[\Delta_0-\kappa\Delta_1\right],\\
\Delta_0&=&\frac{1}{e^{-2\Phi}\sqrt{g}}\p_\mu(e^{-2\Phi}\sqrt{g}g^{\mu\nu }\p_\nu),\quad
\Delta_1=\frac{1}{e^{-2\Phi}\sqrt{h}}\p_\mu(e^{-2\Phi}\sqrt{h}h^{\mu\nu }\p_\nu).\nonumber 
\eea
It is instructive to compare the last expression with the Hamiltonian of the sigma model (\ref{lmbDefrmSigma}) 
\cite{HMSLmb}:
\bea\label{Ham}
\mathcal{H}=-\frac{2\pi}{k} \frac{1+\lambda^2}{1-\lambda^2}\big( \mathcal{L}^\alpha \mathcal{L}^\alpha+\frac{2\lambda}{1+\lambda^2}\mathcal{L}^\alpha \mathcal{R}^\alpha \big)\equiv -\frac{2\pi}{k}\frac{1+\lambda^2}{1-\lambda^2}(\mathcal{H}_0-\kappa\mathcal{H}_1),
\eea
Here $\mathcal{L}$ and $\mathcal{R}$ are the Kac-Moody currents of gauged WZW model. In the undeformed case, the Hamiltonian 
$\mathcal{H}_0$ has been identified with the Laplace operator $(-\Delta_0)$ \cite{Bars2}, and the comparison between (\ref{LapOpe}) and (\ref{Ham}) suggests a similar map between 
$\mathcal{H}_1$ and $(-\Delta_1)$ in the deformed model. Let us now explore the effects of 
$\mathcal{H}_1$ on the undeformed wavefunctions constructed in section \ref{SecGroup}.

In the absence of deformation, the spectrum of the scalar field is determined by the eigenvalues of the Casimir operators for $G$ and $H$ (see (\ref{SfetsosEigen})). The deformation 
\bea
\mathcal{H}_1=\mathcal{L}^\alpha \mathcal{R}^\alpha
\eea
commutes with the Casimir operator $C_2(R)$,\footnote{Recall that $C_2(R)$ can be written either in terms of 
$\mathcal{L}^{M}$ or in terms of $\mathcal{R}^{M}$: 
$C_2(R)=\mathcal{L}^M \mathcal{L}^M=\mathcal{R}^{M}\mathcal{R}^{M}$.} but not necessarily with $C_2(r)$. This implies that the Hamiltonian (\ref{Ham}) can mix only eigenfunctions (\ref{TensorPrescribe}) that correspond to the same values of $(L_1,L_2,L_3)$. Then properties listed on page \pageref{TensorProperties} guarantee that deformed wavefunctions must be polynomials similar to 
(\ref{ConjPrprt1}) and (\ref{ConjPrprt1a}):
\bea\label{PolynDeformed}
{\hat\Psi}&=&\frac{1}{Y^{2p}}P^{(p)}_{m_p;k}+
\frac{1}{Y^{2(p-1)}}P^{(p-1)}_{m_{p-1};k}+\dots +
P^{(0)}_{m_{0};k},\quad p=L_1,\quad m_p=L_2+M\,,\nn
{\tilde\Psi}&=&\frac{\gamma}{Y^{2p}}{\tilde P}^{(p)}_{m_p;k}+
\frac{\gamma}{Y^{2(p-1)}}{\tilde P}^{(p-1)}_{m_{p-1};k}+\dots +
\gamma {\tilde P}^{(0)}_{m_{0};k},\quad \gamma=iabX_3.
\eea
Note that the quantum numbers $(p,m_p)$ do not specify the state completely: even in the undeformed case, $(k,n_p)$ were also 
needed, but they are no longer available since the deformation $\mathcal{H}_1$ does not commute with $C_2(r)$. We will now discuss the structure of mixing between wavefunctions (\ref{ConjPrprt1}), (\ref{ConjPrprt1a}) with different values of $(k,n_p)$. 

\bigskip

Let us  denote a matrix element of the coset in a representation $R$ by $g^R_{MN}$. Then 
the deformation operator $\mathcal{H}_1$ acts on the vector space of $g^R_{MN}$ as
\bea\label{LRaction}
\mathcal{L}^\alpha \mathcal{R}^\alpha g^R_{MN}=T^{R\alpha}_{MP}g^R_{PQ}T_{QN}^{R\alpha} ,
\eea 
where $T^{R\alpha}$ are the generators of the coset in the representation $R$. Since any tensor representation $R$ can be constructed from a product of $q$ copies of the fundamental one, each of the indices $(M,N,P,Q)$ appearing in the last expression can be viewed as a set of $q$ fundamental indices. Thus to understand the structure of $\mathcal{H}_1$, it is sufficient to analyze (\ref{LRaction}) in the fundamental representation. In this case, the generators corresponding to the $SO(6)/SO(5)$ coset are\footnote{Extension to the  $SO(N)/SO(N-1)$, which will be used in sections \ref{SecS3} and \ref{SecSN}, is straightforward: index $n$ takes values from $2$ to $N$.}
\bea 
(T_{1a})_{ij}=\delta_{1i}\delta_{aj}-\delta_{1j}\delta_{ai},~~a=2,\dots,6\,,
\eea 
and their action on the group matrix $g$ is given by
\bea 
\mathcal{L}^\alpha \mathcal{R}^\alpha g_{ij}&=& 
\sum_{a=2}^{6} (T_{1a})_{im}g_{mn}(T_{1a})_{nj}\nn
&=&\sum_{a=2}^{6}(\delta_{1i}\delta_{aj}g_{a1}+\delta_{ia}\delta_{1j}g_{1a}-\delta_{ai}\delta_{aj}g_{11}-\delta_{1i}\delta_{1j}g_{aa}).
\eea 
Thus there are four possible transformations under the deformation $\mathcal{H}_1$:
\bea 
g_{1a}\rightarrow g_{a1},\quad g_{a1}\rightarrow g_{1a},\quad 
g_{11}\rightarrow -\sum_{a=2}^{6}g_{aa},\quad g_{aa}\rightarrow -g_{11}\,.\nonumber
\eea 
Effectively, the deformation removes or adds $1$ in some boxes of the Young tableau. 

For example, in the case of the fundamental representation, only two diagrams will mix, leading to a $2\times 2$ matrix for the deformed $SO(6)/SO(5)$:
\bea\label{Matr100}
\begin{array}{c}
\begin{ytableau}
1
\end{ytableau}\\
\\
\begin{ytableau}
~
\end{ytableau}
\end{array}:
 \quad H^{[1,0,0]}=\begin{bmatrix}
	20&-4\kappa \\
	-20\kappa &4
\end{bmatrix}\,.
\eea
The eigenvalues of the matrix $H^{[1,0,0]}$ give the spectrum of $\Lambda$ in this sector. The undeformed result agrees with (\ref{UndfEigValS5}):
\bea
\Lambda^{[1,0,0]}_{[0,0]}=20,\quad \Lambda^{[1,0,0]}_{[1,0]}=4.
\eea 
Another interesting set of examples is given by a family
\bea\label{ProtectedStates}
L_1=L_2=|L_3|.
\eea
Such states are protected due to the kinematical restriction (\ref{UndfEigValS5}), so the wavefunctions on the deformed geometry must be equal to $\Psi^{[L_1,L_1,L_3]}_{[L_1,L_1]}$, and the eigenvalues are given by
\bea
\Lambda=4(1-\kappa)\left[L_1(L_1+4)+L_1(L_1+2)+L_1^2-L_1(L_1+3)-L_1(L_1+1)\right]\,.\nonumber
\eea
This is a product of $(1-\kappa)$ and (\ref{UndfEigValS5}) with $l_1=l_2=L_1$.

More examples are presented in the Appendix \ref{SecAppEx2}, but it is clear that the group--theoretic method becomes rather cumbersome when the dimension of the representation grows. However, the construction presented in this subsection guarantees that all wavefunctions with given $(L_1,L_2,L_3)$ are linear combinations of polynomials (\ref{PolynDeformed}). Furthermore, the wavefunctions with fixed $L_3$ constructed from (\ref{PolynDeformed}) using the definition (\ref{TensorPrescribe}) satisfy the scaling (\ref{ScaleY})--(\ref{ScaleQ}), which does not rely on values of $(l_1,l_2)$:
\bea\label{ScaleYDeform}
\Psi^{[L_1,L_2,L_3]}&\sim&\frac{1}{Y^{2p}}Q(\alpha,\beta;x,y),\quad
Q(\alpha,\beta;x,y)\sim {\cal A}^{L_2+L_3}.
\eea
These properties allow one to determine the spectrum by applying the rescaling introduced in section \ref{SecUndefLimit} and solving some simple differential equations\footnote{The rescaling modifies the boundary conditions, so the new eigenvalue problem may not be equivalent to the old one. However, combination of the rescaling and the knowledge that the eigenfunction is a polynomial of a certain degree (as proven in this subsection) leads to the desired result.}. We will do this in the next subsection.

\subsection{Scalar on the rescaled geometry}
\label{SecDefResc}

To find the analytic expressions for eigenvalues, we apply the limits (\ref{SeqLimits}) to the deformed geometry (\ref{Simds}). The resulting counterpart of the metric (\ref{UndefGeomLimit}) reads
\bea\label{DeformedGeomLimit}
ds^2&=&\frac{k}{2\pi}\frac{1+\lambda^2}{1-\lambda^2}\left\{-\frac{dY^2}{Y^2}+\frac{dx d\alpha}{2x\alpha}+\frac{dy d\beta}{2y\beta}
-\frac{d\alpha^2}{4\alpha^2}\left[1+\frac{1}{x}+\frac{2y}{x}\right]
-\frac{d\beta^2}{4\beta^2}\left[1+\frac{1}{y}\right]\right\}\nn
&&+\frac{k}{2\pi}\frac{2\lambda}{1-\lambda^2}\left\{\frac{2dXdY}{Y}-\frac{2XdY^2}{Y^2}-\frac{d\alpha dy}{\alpha}-\frac{dy^2}{2y}
-\frac{dx^2}{2x}+\frac{y dx d\alpha}{\alpha x}-\frac{X y d\alpha^2}{2x\alpha^2}\right\}\nn
\eea
Here we introduced a convenient variable $X=x+y$, which will be used below. The metric (\ref{DeformedGeomLimit}) is invariant under independent rescalings of $Y$, $\alpha$, and $\beta$. This implies that polynomial solutions of the Helmholtz equation (\ref{ScalarEqn}) must have the form
\bea\label{PsiCC} 
\Psi=\frac{1}{Y^{2L_1}}\alpha^{L_2}\beta^{L_3}G[X,y],\quad X\equiv x+y.
\eea 
The powers of $(Y,\alpha,\beta)$ in terms of $(L_1,L_2,L_3)$ parameterizing the shape of the Young tableaux were determined in the last subsection. We have also shown that function $G[X,y]$ must be polynomial in both of its arguments. The eigenvalue problem (\ref{ScalarEqn}) for the function 
(\ref{PsiCC}) leads to a differential equation for $G[X,y]$:
\bea\label{FullEqu}
&&X(e_1+e_2X+e_3X^2)\p^2_XG+ y(f_1+f_2y+f_3y^2)\p^2_yG+(c_1+c_2 y+c_3 y^2)\p_y G+\nn
&&\qquad+
y(d_1+d_2 y+d_3 X+d_4 Xy)\p_{Xy}^2G+(b_1+b_2X+b_3y+b_4 X^2+b_5 Xy)\p_X G\nonumber\\
&&\qquad+(a_1+a_2 X+a_3 y)G=0
\eea
Various constants appearing above are given by
\bea
&&a_1=E+4(L_2+L_3)-4L_1(4+L_1)+4\kappa (L_1^2+L_2+L_3),~~a_2=8(L_1-L_2)^2\kappa,\nonumber\\
&&a_3=-8(L_2-L_3)(-1-2L_1+L_2+L_3)\kappa,\nn
&&b_1=d_1=d_2=2c_1=2e_1=2f_1=8(1+\kappa),\\
&&b_2=8[2+L_2+\kappa(1-2L_1+L_2)],~~b_3=8(L_3-L_2)(1+\kappa),~~b_4=8(1-2L_1+2L_2)\kappa,\nonumber\\
&&b_5=16(L_3-L_2)\kappa,~~c_2=8[1+L_3+\kappa(L_3-2L_1)],~~c_3=16(L_3-L_1)\kappa,\nonumber\\
&&d_3=d_4=2e_3=2f_3=16\kappa,~~e_2=f_2=4(1+3\kappa).\nonumber
\eea 
Every polynomial solution of (\ref{FullEqu}) has some highest power $p$ of $X$ and some highest power 
$q$ of $y$, to keep track of these numbers, we write
\bea\label{PolynAnstz}
G=Q^{(p,q)}(X,y).
\eea
Let us now discuss some properties of the polynomials $Q^{(p,q)}(X,y)$.

Isolating the coefficient in front of the highest power of $X$ in $Q^{(p,q)}$,
\bea
Q^{(p,q)}(X,y)=X^p g_p(y)+O(X^{p-1}),
\eea
and looking at the coefficient in front of $X^{p+1}$ in (\ref{FullEqu}), we find a relation
\bea
\kappa(L_1-L_2-p)g_p(y)=0,\nonumber
\eea
This determines the power $p$ in terms of $(L_1,L_2,L_3)$:
\bea\label{DeformedP}
p=L_1-L_2\,.
\eea
Similarly, writing 
\bea
Q^{(p,q)}(X,y)=y^q h_q(X)+O(y^{q-1})\nonumber
\eea
and looking at the coefficient in front of $y^{q+1}$ in (\ref{FullEqu}), we find
\bea\label{DeformedQ}
q=L_2-L_3\,.
\eea

Although equations (\ref{DeformedP}) and (\ref{DeformedQ}) were derived assuming nonzero $\kappa$, it is instructive to compare them with similar relations (\ref{PsiLim2}) and (\ref{LmbLmapUndef}) in the undeformed case. In particular, expression (\ref{PsiLim2}) can be rewritten as
\bea
Q(X,y)=X^\la g_0(y)+X^{\la-1}{g}_1(y)+\dots,
\eea
where $g_0$ is a polynomial of degree $\nu$. According to (\ref{LmbLmapUndef}), $\la=l_1-L_2$ for a fixed value of $l_1$. As we showed in subsection \ref{SecDefLapl}, in the deformed case, function $Q^{(p,q)}(X,y)$ is a mixture of polynomials with all values of $l_1$ allowed by the inequalities (\ref{UndfEigValS5}), and the corresponding powers $\la$ satisfy the relation
\bea
\la=l_1-L_2\le L_1-L_2\,.
\eea
Then $p$ must be equal to the largest allowed value of $\la$, and saturation of the last inequality indeed reproduces (\ref{DeformedP}). Similarly, the second equation in (\ref{LmbLmapUndef}) leads to an inequality for all allowed values of $l_2$:
\bea\label{NuEqnTempA}
\nu=l_2-L_3\le L_2-L_3,
\eea
and saturation reproduces the highest power in the linear combination (\ref{DeformedQ}). This serves as a nice consistency check of the group theoretic construction introduced in section \ref{SecGroup}. Note, however, that, as discussed in the end of section \ref{SecUndefLimit}, only wavefunctions with $\nu\le L_2-|L_3|$ correspond to physical states. In the undeformed case, the easiest way to construct all relevant eigenfunctions involved solving the problem for non--negative values of $L_3$ and using the map (\ref{NuMapL3}), (\ref{PsiLim1mapQQ}). Remarkably, the symmetry (\ref{BetaMapSmtry}) responsible for this simplification is preserved by the deformation (\ref{DeformedGeomLimit}), so the sectors with quantum numbers $(L_1,L_2,L_3)$ and $(L_1,L_2,-L_3)$ have identical spectra. 
Then, without loss of generality, in the rest of this section we will focus only on non--negative values of $L_3$. 

Once the conditions (\ref{DeformedP}) and (\ref{DeformedQ}) are imposed, the left--hand side of equation (\ref{FullEqu}) becomes a polynomial of degree $p$ in $X$ and degree $q$ in $y$. Then (\ref{FullEqu}) reduces to ${\cal N}=(p+1)(q+1)$ homogeneous linear equations for $(p+1)(q+1)$ coefficients of the polynomial $Q^{(p,q)}(X,y)$, and the resulting system has a nontrivial solution if and only if the appropriate matrix (which will be denoted by $M^{[L_1,L_2,L_3]}$) has vanishing determinant. This leads to a characteristic equation of degree ${\cal N}$ for $\Lambda$, which always has ${\cal N}$ solutions. For example, $(L_1,L_2,L_3)=(1,0,0)$ has $p=1$, $q=0$, and $\Lambda$ is an eigenvalue of the $2\times 2$ matrix (\ref{Matr100}). Thus the question of finding the eigenvalues of the Helmholtz equation (\ref{ScalarEqn}) is reduced to a straightforward linear algebra problem, and the next subsection is devoted to the detailed analysis of matrices $M^{[L_1,L_2,L_3]}$ and their eigenvalues.

\subsection{Spectrum from algebraic equations}
\label{SecDefMatr}

Let us determine the structure of the matrices $M^{[L_1,L_2,L_3]}$ and explore the properties of their eigenvalues. It is convenient to specify the $(L_1,L_2,L_3)$ sector by $(L_1,p,q)$, where $(p,q)$ are given by (\ref{DeformedP}) and (\ref{DeformedQ}):
\bea\label{DeformedPQ}
p=L_1-L_2,\quad q=L_2-L_3\,.
\eea
As discussed after equation (\ref{NuEqnTempA}), to eliminate unphysical states we focus only on non--negative values of $L_3$,\footnote{To find the full spectrum, one should to assign double degeneracy to the eigenvalues $\Lambda$ corresponding $L_3\ne 0$. This amounts to including states with both $L_3$ and $(-L_3)$, whose wavefunction are mapped into each other by the coordinate transformation (\ref{BetaMapSmtry}) that leaves the metric (\ref{DeformedGeomLimit}) invariant.} then $p$ and $q$ are arbitrary non--negative integers subject to one constraint:
\bea
p+q\ge 0.
\eea
To derive the matrix elements, we start with a polynomial ansatz
\bea 
G=\sum_{m}^{p}\sum_{n=0}^{q}c_{mn} X^m y^n,
\eea 
and substitute it  into (\ref{FullEqu}). This leads to a system of linear equations with constant coefficients: 
\bea\label{LinearSystsS5} 
c_{mn} C_{mn}^{00}+c_{m+1,n}C_{mn}^{10}+c_{m,n+1}C_{mn}^{01}+c_{m+1,n-1}C_{mn}^{1\bar{1}}
+c_{m-1,n}C_{mn}^{\bar{1}0}+c_{m,n-1}C_{mn}^{0\bar{1}}=0.
\eea 
Here we introduced a set of numbers:
\bea
&&C_{mn}^{00}=\Lambda-4(1-\kappa)(L_1+1-m-n)^2+8\nonumber\\
&&\qquad+4(1+\kappa)\left[2 (m^2+n^2+mn)-m (2 p-1)-n (2 p+2 q+1)-2 p-q-1\right],\nonumber\\
&&C_{mn}^{10}=4 (\kappa +1) (m+1) (m+2 n+2),\nonumber\\
&&C_{mn}^{01}=4 (\kappa +1) (n+1)^2, \\
&&C_{mn}^{1\bar{1}}=8 (\kappa +1) (m+1) (n-q-1),\nonumber \\
&&C_{mn}^{\bar{1}0}=8 \kappa  (-m+p+1)^2,\nonumber\\
&&C_{mn}^{0\bar{1}}=8 \kappa   (n-q-1) (2 m+n-2 p-q-2).\nonumber
\eea 
As expected, there are $(p+1)(q+1)$ homogeneous linear equations for $c_{mn}$.  To simplify the form of the corresponding matrix, 
it is convenient to order the variables as 
\bea
\{c_{00},c_{01},\dots,c_{0q};c_{10},c_{11},\dots,c_{1q},\dots\}\,.\nonumber
\eea
With this convention, the matrix describing the system (\ref{LinearSystsS5}) can be written in terms of 
$(q+1)\times (q+1)$--blocks:
\be\label{MatrPQ}
M^{[L_1,L_2,L_3]}=\begin{bmatrix}
	\bold{C}_0 & \bold{B}_{0} &0&0 &\dots\\
	\bold{D}_{1}& \bold{C}_1&\bold{B}_{1}&0&\dots\\
	0&\bold{D}_{2}&\bold{C}_2 & \bold{B}_{2}&\dots \\
	\vdots &\vdots &\ddots &\ddots &\ddots\\
	0&0&0 &\bold{D}_{p} &\bold{C}_p
\end{bmatrix}
\ee 
Using the freedom of scaling and shifting  $X$ and $y$, one can cancel the coefficients $C_{mn}^{1\bar{1}}$. Then, remarkably, both $B_{m}$ and $D_{m}$ transform into constant diagonal matrices. Specifically, changing $(X,y)$ as
\be 
X\rightarrow 2\kappa X+1+\kappa ,\quad y\rightarrow y,
\ee 
we find
\bea\label{MatrPQa} 
&&\bold{B}_m=4 \left(\kappa ^2-1\right) (m+1) (2 L_1-m+3) \bold{I}_{(q+1)\times(q+1)},\nn
&&\bold{D}_{m}=4(1-m+p)^2\bold{I}_{(q+1)\times(q+1)}.
\eea 
The central blocks ${\bf C}_m$ also simplify:
\bea\label{MatrPQb} 
{\bf C}_m=\begin{bmatrix}
	C_{m0}^{00} & C_{m0}^{01}&0&0 &\dots\\
	C_{m1}^{0\bar{1}}& C_{m1}^{00} &C_{m1}^{01}&0&\dots\\
	0&C_{m2}^{0\bar{1}}&C_{m2}^{00} & C_{m2}^{01}&\dots \\
	\vdots &\vdots &\ddots &\ddots &\ddots\\
	0&0&0 &C_{mq}^{0\bar{1}} &C_{mq}^{00}
\end{bmatrix}_{(q+1)\times(q+1)}	
\eea 
Here
\bea\label{MatrPQc} 
&&C_{mn}^{00}=\Lambda-4(1-\kappa)\left[(L_1-m-n+1)^2-1\right]\nonumber\\
&&\quad \quad +4(1+\kappa)[2n^2+2mn-m^2+2m(1+p)-n(1+2p+2q)-p^2-2p-q].\nonumber\\
&&C_{mn}^{01}=4 (1+\kappa) (n+1)^2,\\
&&C_{mn}^{0\bar{1}}=4 \kappa  (n-q-1) (2 m+n-2 p-q-2),\nonumber
\eea 
To summarize, we have arrived at a simple {\it algebraic} procedure for constructing the eigenvalues of the {\it differential equation} (\ref{ScalarEqn}):
\label{StepsS5}
\begin{enumerate}[(i)]
\item Start with a set of $SO(6)$ quantum numbers $(L_1,L_2,L_3)$ and define the parameters $(p,q)$ using (\ref{DeformedPQ}).
\item Construct the corresponding matrix 
$M^{[L_1,L_2,L_3]}$ using (\ref{MatrPQ}), (\ref{MatrPQa}),  (\ref{MatrPQb}),  (\ref{MatrPQc}). 
\item Eigenvalues $\Lambda$ in a particular $(L_1,L_2,L_3)$ sector are determined by solving equation
\bea\label{detMS5}
\mbox{det}\, M^{[L_1,L_2,L_3]}=0.
\eea
The left--hand side is a polynomial of degree $(p+1)(q+1)$ in $\Lambda$. 
\end{enumerate}
Several examples of matrices $M^{[L_1,L_2,L_3]}$ and the resulting polynomials are presented in the Appendix \ref{SecAppEx2}.

\section{Scalar field on the deformed $SO(4)/SO(3)$ coset}
\label{SecS3}
\renewcommand{\theequation}{5.\arabic{equation}}
\setcounter{equation}{0}

While we are mostly interested in the spectrum of the Helmholtz equation (\ref{ScalarEqn}) on the deformed five--sphere\footnote{Strictly speaking, the metric (\ref{Simds})--(\ref{Simds1}) is a lambda--deformation of the non--abelian T--dual of $S^5$, but we refer to it as a deformed five--sphere since this shortcut is used in the literature. Similarly, we will refer to the lambda--deformation of the $SO(k+1)/SO(k)$ coset as a deformed $S^k$.\label{foot11}}, and all relevant results were presented in the last section, it might also be interesting to look at a deformed $S^3$. Let us briefly discuss the properties of this geometry. 

\bigskip

Our starting point is the gauged WZW model for $SO(4)/SO(3)$. To construct the spectrum of the scalar field on this geometry we follow the logic outlined in section \ref{SecGroup}. 
Representations of $SO(4)$ are characterized by two numbers $(L_1,L_2)$, which can be either integers or half--integers, while representations of $SO(3)$ are characterized by one such number. Then in the geometric limit of large $k$, the general relation (\ref{SfetsosEigen}) gives
\bea\label{UndfEigVal43}
SO(4)/SO(3):\  \Lambda=4\Big[L_1(L_1+2)+L_2^2-l_1(l_1+1)\Big]\,,\quad
L_1\ge l_1\ge |L_2|.
\eea
We will denote the corresponding wavefunction by
\bea\label{UndfEigFnc43}
\Psi^{[L_1,L_2]}_{[l_1]}\,.
\eea

We parameterize the $SO(4)/SO(3)$ coset using a low--dimensional version of (\ref{CosetG1}) and impose an obvious counterpart of the gauge condition (\ref{CosetG3}):
\bea 
&&A=\begin{bmatrix}
	0&0&0\\
	0&0&a\\
	0&-a&0
\end{bmatrix},\quad X=(X_2,X_1,0)\,.
\eea 
The deformed metric is still given by (\ref{Simds}), but now $J$ is a $3\times 3$ matrix,
\be
J=\begin{bmatrix}
	1&0\\
	0&W(a)
\end{bmatrix},\quad W(x)=\begin{bmatrix}
	\frac{x^2-1}{x^2+1}&\frac{2x}{x^2+1}\\
	\frac{2x}{x^2+1}&\frac{1-x^2}{x^2+1}
\end{bmatrix}\,,
\ee
and the frames $e_{(0)}$ are given by 
\bea 
&&e_{(0)}^1=\frac{da X_2}{a(1+a^2)}+\frac{dX_2+dX_2 X_1^2-dX_1 X_1 X_2}{Y^2},\nn
&&e_{(0)}^2=dX_1-\frac{da X_2^2}{a X_1(1+a^2)}-\frac{X_1(dX_1+X_2dX_2)}{Y^2},\\
&&e_{(0)}^3=-\frac{da}{X_1(1+a^2)},\quad Y^2\equiv1+X_1^2+X_2^2.\nonumber
\eea 
The discussion of section \ref{SecGroup} can be repeated after introduction of coordinates $(\alpha,x)$ 
defined by (\ref{DefineABY}), and now we summarize the results.

\bigskip
\noindent
{\bf Properties of the eigenfunction $\Psi^{[L_1,L_2]}_{[l_1]}$ for the tensor representations:}
\begin{enumerate}
\item 
The wavefunction $\Psi^{[L_1,L_2]}_{[l_1]}$ corresponding to a given representation of the 
$SO(4)/SO(3)$ coset is given by a sum of two expressions:
\bea
&&\Psi^{[L_1,\pm M]}_{[l_2]}={\hat\Psi}^{[L_1,M]}_{[l_1]}\pm 
{\tilde \Psi}^{[L_1,M]}_{[l_1]}\,,\quad M=|L_2|\,.
\eea
Here ${\hat\Psi}$ and ${\tilde\Psi}$ are polynomials of 
degree $p=L_1$ in $Y^{-2}$:
\label{PagePropS3}
\bea\label{PsiForS3}
{\hat\Psi}&=&\frac{1}{Y^{2p}}P^{(p)}_{m_p,n_p;k}+
\frac{1}{Y^{2(p-1)}}P^{(p-1)}_{m_{p-1},n_{p-1};k}+\dots +
P^{(0)}_{m_{0},n_{0};k},\\
{\tilde\Psi}&=&\frac{\gamma}{Y^{2p}}{\tilde P}^{(p)}_{m_p,{\tilde n}_p;k}+
\frac{\gamma}{Y^{2(p-1)}}{\tilde P}^{(p-1)}_{m_{p-1},{\tilde n}_{p-1};k}+\dots +
\gamma {\tilde P}^{(0)}_{m_{0},{\tilde n}_{0};k},\quad \gamma=aX_2\,.\nonumber
\eea 
\item Expressions $P^{(q)}_{m_q,n_q;k}$ and ${\tilde P}^{(q)}_{m_q,n_q;k}$ are polynomials in 
$(\alpha,x)$ that have degree $m_q$ in $\alpha$, degree $n_q$ in $x$, and degree $k$ in $(x,\alpha)$.
The powers in the leading polynomials 
$(P^{(p)}_{m_p,n_p;k},{\tilde P}^{(p)}_{m_p,{\tilde n}_p;k})$ are given by 
\bea
&&p=L_1,\quad k=l_1,\quad m_p=L_2+M,\quad n_p=l_1,\quad {\tilde n}_p=l_1-1\,.
\eea
\item For large values of $\alpha$ and fixed $x$, two useful combinations of expressions (\ref{PsiForS3}) scale as
\bea
\Big[P^{(p)}_{m_p,n_p;k}-\gamma{\tilde P}^{(p)}_{m_p,{\tilde n}_p;k}\Big]_{Y=0}\sim \alpha^{-M},\quad
\Big[P^{(p)}_{m_p,n_p;k}+\gamma{\tilde P}^{(p)}_{m_p,{\tilde n}_p;k}\Big]_{Y=0}\sim \alpha^{M}\,.
\nonumber
\eea 
\item The three properties listed above imply that in the $Y=0$ limit, the wavefunction can be written as
\bea\label{S3CountProp}
\Psi^{[L_1,L_2]}_{[l_1]}&\sim&\frac{1}{Y^{2L_1}}\left[Q(\alpha,x)+O(Y^2)\right],\quad
Q(\alpha,x)=\alpha^{L_2}\left[{\tilde Q}(x)+O(\alpha^{-1})\right]
\eea
This scaling will play a crucial role in our study of eigenfunctions and eigenvalues.
\item Although the expressions $(P^{(p)}_{m_p,n_p;k},{\tilde P}^{(p)}_{m_p,n_p;k})$ with $q<p$ will not be needed for our discussion, we note that as $q$ decreases, only the following options are 
possible:
\bea
n_{q-1}=\{n_q,n_{q}-1\},\quad m_{q-1}=\{m_q,m_{q}+1\}
\eea
\end{enumerate}
As in the $SO(6)/SO(5)$ case discussed in section \ref{SecUndeform}, the $SO(4)/SO(3)$ coset also admits the ``spinor representations'', which are not described by the Young tableaux, and which correspond to half--integer values of $(L_1,L_2,l_1)$. The counterparts of the elementary building blocks (\ref{PsiHalf}) are
\bea\label{PsiHalfS3}
\Psi^{[\frac{1}{2},\frac{1}{2}]}_{[\frac{1}{2}]}=\frac{\sqrt{\alpha}}{Y}
[1+\gamma],\quad
\Psi^{[\frac{1}{2},-\frac{1}{2}]}_{[\frac{1}{2}]}=\frac{\sqrt{\alpha}}{Y}
[1-\gamma],\quad \Lambda=\frac{3}{2}.
\eea
Clearly these wavefunctions satisfy the counting (\ref{S3CountProp}). An arbitrary representation must have either integer of half--integer values of $(L_1,L_2,l_1)$, and all wavefunctions can be constructed using the logic that led to (\ref{HalfFusion}). In particular, all eigenfunctions satisfy the scaling (\ref{S3CountProp}). 

\bigskip

After establishing the five properties listed above, we can determine the eigenvalues using the rescaling similar to (\ref{SeqLimits}):
\bea\label{SeqLimitsS3}
Y\rightarrow 0,\ \mbox{then}\ \alpha\rightarrow\infty,\ 
\mbox{fixed}\ x.
\eea
To shorten the discussion we apply the scaling (\ref{SeqLimitsS3}) directly to the deformed geometry. This leads to the metric and the dilaton
\bea\label{DefS3metrResc}
ds^2&=&\frac{k}{2\pi}\frac{1+\lambda^2}{1-\lambda^2}
\Big[-\frac{dY^2}{Y^2}-\frac{d\alpha^2}{4x\alpha^2}(1+x)+\frac{d\alpha dx}{2x\alpha } 
\Big]\\
&& +\frac{k}{2\pi}\frac{2\lambda}{1-\lambda^2} \Big[-\frac{(1+2x)dY^2}{Y^2}-\frac{d\alpha^2}{4x\alpha^2}(1+x)+\frac{d\alpha dx}{2x\alpha }-\frac{dx^2}{2x}+\frac{2dxdY}{Y}\Big],\nn
e^{-2\Phi}&=&\frac{x}{Y^4}\,.\nonumber
\eea 
Symmetries of these expressions imply that equation (\ref{ScalarEqn}) is invariant under rescalings of $Y$ and $\alpha$, so the eigenfunctions are simple powers of $(Y,\alpha)$. Then (\ref{S3CountProp}) relates the resulting exponents with the shape of the Young tableau\footnote{Recall that for tensor representations, $L_1$ and $|L_2|$ represent the lengths of the rows of the Young tableau.}:
\bea\label{PsiS3Anstz} 
\Psi^{[L_1,L_2]}_{l_1}=\frac{1}{Y^{2L_1}}\alpha^{L_2}Q(-x)\,.
\eea 
Substitution of the last expression into the field equation (\ref{ScalarEqn}) with the metric (\ref{DefS3metrResc}) leads to a second order ODE for the {\it polynomial} $Q(z)$:
\bea\label{HeunE}
Q''(z)+\big[\frac{1}{z}+\frac{1+2L_2}{z-1}-\frac{1+2L_1}{z-c}\big]Q'(z)+
\frac{(L_1-L_2)^2(z-h)}{z(z-1)(z-c)}Q(z)=0\,.
\eea
This is the well-known Heun's equation \cite{Heun}\footnote{Note that we wrote $Q(-x)$ in the ansatz (\ref{PsiS3Anstz}) to ensure that (\ref{HeunE}) reproduces the standard form of the Heun's equation for $Q(z)$.}, and its parameters $(c,h)$ are given by
\bea 
c=\frac{1+\kappa}{2\kappa},\quad 
h=\frac{(1-2c)L_1+(1-c)L_1^2+cL_2 }{\left(L_1-L_2\right){}^2}+
\frac{\Lambda}{8\kappa (L_1-L_2)^2}.
\eea 
Equation (\ref{FullEqu}) encountered for the deformed $S^5$ can be viewed as a counterpart of the Heun's equation for functions of two variables.

Let us recall the procedure for finding polynomial solutions of the Heun's equation. 
Starting with the ansatz $Q_p(z)$ that truncates at power $p$,
\bea\label{HeunPolyn}
Q_p(z)=\sum_{n=0}^p a_n z^n,\quad a_p\ne 0,
\eea 
one arrives at the recursion relation for the coefficients:
\bea\label{HeunRecur} 
&&a_{n+1}=A_n a_n+B_n a_{n-1},\nonumber \\
&&A_n=\frac{(1+c)n^2+[(1+2L_2)c-(1+2L_1)]n}{c(n+1)^2}+
\frac{(L_1-L_2)^2 h}{c(n+1)^2},\\
&&B_n=-\frac{(n-1+L_2-L_1)^2}{c(n+1)^2}\nonumber
\eea 
Then the truncation condition (\ref{HeunPolyn}) implies that 
\bea 
p=L_1-L_2
\eea 
and
\bea\label{detS3}
\mbox{det}(M_{p+1})=0,\ \mbox{where}\  M_{p+1}=\begin{bmatrix}
	A_0&-1&0&0&0&\dots &\dots &\dots\\
	B_1&A_1&-1&0&0&\dots &\dots &\dots\\
	0&B_2&A_2&-1&0&\dots &\dots &\dots\\
	\vdots & \vdots & \vdots & \vdots & \vdots &\dots &\vdots &\vdots &\\
	\dots & \dots &\dots &\dots &\dots & \dots & A_{p-1}&-1\\
	0&	0&	0&	0&	0&\dots & B_p& A_p
	\end{bmatrix}\,.
\eea 
As in the $SO(6)/SO(5)$ case, we find that for positive values of $L_2$, all polynomial solutions found here represent limits of eigenfunctions on the full deformed coset. In contrast to this, some polynomial solutions with negative $L_2$ are just artifacts of the limit, so the eigenvalues corresponding to them do not contribute to the physical spectrum. We refer to the end of section \ref{SecUndefLimit} for the detailed discussion of this point, and here we just formulate the result adjusted to the $SO(4)/SO(3)$ coset. To find the physical eigenvalues, one solves equation (\ref{detS3}) only in sectors with $L_2\ge 0$. The subspaces with negative $L_2$ affect only the degeneracies: every value of $\Lambda$ with non--zero $L_2$ should be counted twice. The physical eigenfunctions with $(L_1,L_2)$ and $(L_1,-L_2)$ are mapped into each other by a counterpart of the coordinate transformation (\ref{BetaMapSmtry})
\bea
\alpha\rightarrow \frac{(1+x)^2}{\alpha}\,,
\eea
which leaves the metric (\ref{DefS3metrResc}) invariant. 

To summarize, we conclude that any pair of (half--)integers $(L_1,L_2)$ that satisfy an inequality
\bea
L_1\ge |L_2|,
\eea
gives rise to $(L_1-|L_2|+1)$ eigenvalues of the Helmholtz' differential equation (\ref{ScalarEqn}). 
For non--negative $L_2$, the values  of $\Lambda$ are determined by solving an {\it algebraic} equation (\ref{detS3}) of degree $(L_1-L_2+1)$. The contributions from negative $L_2$ are added by observing that the sectors with $(L_1,L_2)$ and $(L_1,-L_2)$ lead to the same {physical} eigenvalues. Some explicit examples are presented in the Appendix \ref{SecAppEx3}.

We conclude this section by recovering the undeformed eigenvalues (\ref{UndfEigVal43}). For $\kappa=0$, the number of poles in the Heun's equation (\ref{HeunE}) reduces to two, and the result reads
\bea
z(z-1)Q''(z)-\left[1-2(1+L_2)z\right]Q'(z)+\left[\frac{\Lambda}{4}-L_1(L_1+2)+L_2\right]Q(z)=0
\eea
Only one solution of this hypergeometric equation is regular at $z=0$,
\bea
&&Q(z)=F\left[-\nu,1+2L_2+\nu,1;z\right],\nn 
&&\Lambda=4(L_1+1)^2+4L_2^2-(2\nu+2L_2+1)^2-3.
\eea
and $Q(z)$ becomes a polynomial if $\nu$ is a non--negative integer. The expected result (\ref{UndfEigVal43}) is reproduced upon identification $\nu=l_1-L_2$. The parameter $\nu$ 
is indeed a non--negative integer due to the inequality (\ref{UndfEigVal43}). 

Rather than going through the hypergeometric equation, one can also determine the undeformed spectrum by taking $\kappa$ to zero (or $c$ to infinity) in the Heun's recurrent relations (\ref{HeunRecur}):
\bea\label{HeunRecurHyper} 
a_{n+1}=A_n a_n,\quad
A_n=\frac{n^2+(1+2L_2)n-2L_1-L_1^2+L_2-\frac{\Lambda}{4}}{(n+1)^2}
\eea 
The solution becomes a polynomial of degree $p$ if and only if $A_p=0$, then
\bea
\Lambda=4(L_1+1)^2-4-4p^2-4(1+2L_2)p-4L_2
\eea
Again, one recovers the expected spectrum (\ref{UndfEigVal43}) on the undeformed $S^3$ upon identification $p=l_1-L_2$.

\section{Extensions to other cosets}
\label{SecSN}
\renewcommand{\theequation}{6.\arabic{equation}}
\setcounter{equation}{0}

Although from the string theory prospective, the integrable lambda--deformations are interesting only for $S^3$ and $S^5$, the construction presented in this article can be formally extended to arbitrary odd--dimensional spheres\footnote{See footnote on page \pageref{foot11}.}, and in this short section we briefly discuss such extensions.

\bigskip

We begin with analyzing the spectrum of the Helmholtz equation (\ref{ScalarEqn}) on the undeformed coset $SO(2n)/SO(2n-1)$. The eigenvalues are given by the general expression (\ref{SfetsosEigen}) discovered in \cite{PS}, and to write $\Lambda$ in a more explicit form, we recall that  
representations of $SO(2n)$ are characterized by $n$ numbers $(L_1,\dots L_n)$, which can be either integers or half--integers. Representations of $SO(2n-1)$ are characterized by $(n-1)$ such numbers, 
$(l_1,\dots,l_{n-1})$, and the embedding into $SO(2n)$ implies a set of inequalities \cite{Ma}:
\bea\label{SetOfLs}
 L_1\ge l_1\ge L_2\ge \dots \ge l_{n-1}\ge |L_n|\,.
\eea
For tensor representations, which are specified by integer values of $(L_k,l_k)$, one can interpret 
$(L_1,\dots ,|L_n|)$ as the lengths of the rows in the appropriate Young tableau\footnote{Unless $L_n=0$, a given Young tableau describes two representations, which are dual to each other, and which differ by the sign of $L_n$. We refer to the Appendix \ref{SecAppDual} for the detailed discussion of the duality.}, and 
$(L_1-l_1,\dots ,L_{n-1}-l_{n-1})$ as numbers of the label $1$ is such rows (see the discussion leading to the examples (\ref{ExamplesMarked})). 

The Casimir operators entering (\ref{SfetsosEigen}) can be expressed in terms of $(L_1,\dots L_n,l_1,\dots,l_{n-1})$, and in the geometric limit of large $k$, the eigenvalues become
\bea\label{UndfEigValNN}
\frac{G}{H}=\frac{SO(2n)}{SO(2n-1)}:\quad  \Lambda=4\Big[\sum_{k=1}^n L_k(L_k+2n-2k)-
\sum_{k=1}^{n-1} l_k(l_k+2n-2k-1)\Big]\,.
\eea
Following the notation adopted for $n=(2,3)$, we denote the corresponding wavefunction by
\bea\label{UndfEigFncNN}
\Psi^{[L_1,\dots,L_n]}_{[l_1,\dots l_{n-1}]}\,.
\eea
Note that the eigenfunctions (\ref{UndfEigFncNN}) for the special values of parameters 
\bea
L_1=1, \quad L_2=\dots=L_n=l_n=\dots =l_{n-1}=0 \nonumber
\eea
were used in \cite{GrigTseytl} to get important insights into integrable structures associated with gWZW models. The solutions of \cite{GrigTseytl} corresponded to eigenvalues $\Lambda=4(2n-1)$. Here we are interested in the general set of quantum numbers (\ref{SetOfLs}).

Parameterization of the $G/H$ coset is obtained by a straightforward generalization of 
(\ref{CosetG1})--(\ref{CosetG3}), and  the appropriate coordinates are 
$(a_1,\dots a_{n-1},X_1,\dots X_{n})$. We further define the counterparts of variables (\ref{DefineABY}):
\bea\label{DefineABYNN}
\alpha_k=\frac{1}{1+a_k^2},\quad 
x_k=\frac{a_k^2X_k^2}{1+a_k^2},\quad
Y=\left[1+\sum_{k=1}^{N}X_k^2\right]^{1/2}\,.
\eea
As in the cases of $n=(2,3)$ analyzed in previous sections, one can show that the wavefunctions describing tensor representations are polynomials in $(a_k,x_k,\gamma)$, where 
\bea
\gamma=iX_{n}\prod_{k=1}^{n-1} a_k\,.\nonumber
\eea
We will be particularly interested in the generalization of the limits (\ref{SeqLimits}), (\ref{SeqLimitsS3}):
\bea\label{SeqLimitsNN}
Y\rightarrow 0,\ \mbox{then}\ \alpha_1\rightarrow\infty,\ \mbox{then}\ \alpha_2\rightarrow\infty,\ \dots,\ 
\mbox{then}\ \alpha_{N-1}\rightarrow\infty,\quad 
\mbox{fixed}\ x_k.
\eea
Then the wavefunction (\ref{UndfEigFncNN}) takes the form 
\bea\label{PsiCCNN} 
\Psi^{[L_1,\dots,L_n]}_{[l_1,\dots l_{n-1}]}=\frac{1}{Y^{2L_1}}\alpha_1^{L_2}\alpha_2^{L_3}\dots \alpha_{n-1}^{L_n}G[x_1\dots x_{n-1}].
\eea 
One can further determine the powers of all arguments in the polynomial $G$ in terms of 
$(l_1,\dots,l_{n-1})$, as we did for $S^5$ and $S^3$ in the earlier sections. We will not discuss this further since our main goal is the study of the lambda deformation, which mixes polynomials with all allowed values of $(l_1,\dots,l_{n-1})$. 

\bigskip

Substitution of the ansatz (\ref{PsiCCNN}) into the Helmholtz equation (\ref{ScalarEqn}) for the deformed 
$SO(2n)/SO(2n-1)$ coset leads to a PDE for the {\it polynomial} $G$. The result is a higher--dimensional counterpart of the Heun's equation (\ref{HeunE}) and its generalization (\ref{FullEqu}). Following the pattern found for $S^3$ and $S^5$, we introduce new variables
\bea
y_1=x_1+\dots+x_{N-1},\quad y_2=x_2+\dots+x_{N-1},\quad\dots,\quad y_{N-1}=x_{N-1}\,,
\eea
and assume that the polynomial $G$ has degrees $(p_1,\dots,p_{N-1})$ in $(y_1,\dots,y_{N-1})$: 
\bea\label{GfuncNN}
G=\sum_{k_1=0}^{p_1}\sum_{k_2=0}^{p_2}\dots \sum_{k_{N-1}=0}^{p_{N-1}}
c_{k_1\dots k_{N-1}}y_1^{k_1}\dots y_{N-1}^{k_{N-1}}\,.
\eea
The number of undetermined coefficients is given by
\bea\label{CalCNN}
{\cal C}=\prod_k (p_k+1).
\eea
The consistency conditions for the equation (\ref{ScalarEqn})  lead to generalizations of relations (\ref{DeformedP}), (\ref{DeformedQ}):
\bea\label{DeformedPNN}
p_1=L_1-L_2,\quad p_2=L_2-L_3,\quad\dots\quad p_{n-1}=L_{n-1}-L_n
\eea
With these powers, the left--hand side of the Helmholtz equation (\ref{ScalarEqn}) becomes a polynomial of degrees $(p_1,\dots,p_{n-1})$ in $(y_1,\dots,y_{n-1})$. Then relation (\ref{ScalarEqn}) reduces to a homogeneous system of ${\cal C}$ linear equations for ${\cal C}$ variables 
$c_{k_1\dots k_{n-1}}$. The determinant of the corresponding square matrix must vanish, leading to an algebraic equation of degree ${\cal C}$ for the eigenvalue $\Lambda$. This equation generalizes (\ref{detMS5}) and (\ref{detS3}).

\bigskip
\noindent
We conclude this section by summarizing the procedure for finding the eigenvalues of the Helmholtz equation (\ref{ScalarEqn}) that can be applied to any deformed $SO(2n)/SO(2n-1)$ coset. 
\begin{enumerate}[(i)]
\item Start with a set of $SO(2n)$ quantum numbers $(L_1,\dots,L_n)$, which are either all integers or all half--integers, and determine the corresponding parameters $p_k$ using (\ref{DeformedPNN}).
\item Substitute the ansatz (\ref{PsiCCNN}), (\ref{GfuncNN}) into the Helmholtz equation (\ref{ScalarEqn}), and extract the ${\cal C}\times {\cal C}$ matrix $M^{[L_1,\dots,L_n]}$ describing the resulting homogeneous system of algebraic equations\footnote{As in the $n=2,3$ cases, which were discussed in detail in previous sections, one finds that for negative values of $L_n$ not all ${\cal C}$ states can be viewed as limits of eigenfunctions. To cure this problem, one should focus only on sectors with $L_n\ge 0$ and observe that subspaces with $(L_1,\dots L_{n-1},L_n)$ and $(L_1,\dots L_{n-1},-L_n)$ lead to the same physical eigenvalues.}. 
\item Determine the eigenvalues $\Lambda$ in the $(L_1,\dots,L_n)$ sector by solving an algebraic equation
\bea\label{detMSN}
\mbox{det}\, M^{[L_1,\dots,L_n]}=0.
\eea
The left--hand side is a polynomial in $\Lambda$, and its degree ${\cal C}$ is given by (\ref{CalCNN}). 
\end{enumerate}
Group--theoretic arguments presented in section \ref{SecDeformed} guarantee that this procedure recovers all eigenvalues of the Helmholtz equation (\ref{ScalarEqn}). Note that our construction replaces a complicated differential equation by a simple algebraic problem.

\section{Discussion} 

In this paper we have proposed a new method for finding spectra of scalar fields on a large class of backgrounds without isometries. We extended the construction introduced in article \cite{PS}, which served as the main inspiration for our work, from gauged WZW models on cosets to their integrable deformations. While none of these backgrounds posses geometric isometries, the gauged WZW systems have rich symmetries at the level of the sigma model, and these structures are destroyed by the deformations. 
As the result of this, the group--theoretic procedure of \cite{PS} is not sufficiently powerful to determine the spectrum, so we had to combine it with some insights from studying the structures associated with the deformation and with the differential equation for the scalar field. Remarkably, we were able to find the spectrum of the Helmholtz equation by solving simple {\it algebraic} equations.

In this paper we focused on integrable deformations of the gauged WZW models only because such solutions of string theory are known explicitly, but it is not clear whether integrability has played any role in the success of our construction. It would be very interesting to see whether the methods developed here are  applicable to scalar fields on non--integrable deformations of the gauged WZW models, but unfortunately construction of such solutions of string theory or supergravity is challenging due to lack of isometries. It would also be very interesting to see whether our methods can be extended to vector fields and other excitations.


\section*{Acknowledgments}

This work was supported in part by the DOE grant DE\,-\,SC0017962.

\appendix

\section{Properties of tensor representations}
\label{SecAppA}
\renewcommand{\theequation}{A.\arabic{equation}}
\setcounter{equation}{0}

In this appendix we justify the seven properties of wavefunctions for tensor representations listed on page \pageref{TensorProperties}. In section \ref{SecAppYoung} we establish the map between the Young tableaux and wavefunctions, then in section \ref{SecAppCount} we perform the power counting for function ${\hat\Psi}$ that justifies its properties listed on page \pageref{TensorProperties}. In section \ref{SecAppDual} these results are extended to function ${\tilde\Psi}$ involving ``dual representations'' and to the full wavefunction $\Psi^{[L_1,L_2,L_3]}_{l_1,l_2}$.

\subsection{Wavefunctions and Young tableaux}
\label{SecAppYoung}

In this subsection we provide some details of the construction (\ref{TensorPrescribe}). We begin with reviewing the correspondence between $SO(6)$ tensors and Young tableaux, paying special attention to transformation of such tensors under the subgroup $SO(5)$. In the second part of this subsection these group--theoretic properties will be used to construct the wavefunctions for the $SO(6)/SO(5)$ coset.

\bigskip

An irreducible tensor representation of $SO(6)$ is characterized by a Young tableau with three rows, and we use integers $(L_1,L_2,|L_3|)$ to denote the number of boxes in these rows. The total number of boxes will be denoted by $L$ ($L=L_1+L_2+|L_3|$). An element of a given representation can be written in terms of the basis tensors 
${\theta}_{a_1\dots a_L}$,
\bea\label{CombinedTheta}
\left[{\theta}_{a_1\dots a_L}\right]^{i_1\dots i_L}=\delta_{a_1}^{i_1}\dots \delta_{a_L}^{i_L}\,,
\eea
by performing antisymmetrization of indices within every column, followed by symmetrization within every row. Application of this prescription to representations of $SU(3)$ would have led to irreducible 
tensors\footnote{Recall that for $SO(6)$ the indices in (\ref{CombinedTheta}) take six possible values, while for $SU(3)$ there are only three options.}, but since action of $SO(6)$ commutes with contraction of indices, one encounters a further reduction into the traceless part and the trace. Some examples will be given below. 

We will be interested in transformations of the irreducible $SO(6)$--tensors $S_{a_1\dots a_L}$ under the $SO(5)$ rotations that leave $a=1$ invariant. Using Greek letters\footnote{In this subsection we denote the $SO(6)$ indices by $(a,b,c,\dots)$ and the $SO(5)$ indices by $(\alpha,\beta,\gamma,\dots)$. The rest of this article uses only the $SO(5)$ indices which are denoted by $(a,b,c\dots)$. This change of notation makes the formulas cleaner, and we hope that it does not lead to confusions.}  to denote $a\ne 1$, we conclude that $S_{1\dots 1}$ is a scalar, $S_{\alpha 1\dots 1}$ is a vector, and so on. However, 
$S_{\alpha\beta 1\dots 1}$ is a reducible tensor since it contains a contribution from trace.  
The irreducible object is
\bea
T_{\alpha\beta 1\dots 1}\equiv S_{\alpha\beta 1\dots 1}-\frac{1}{5}\delta_{\alpha\beta}
\delta^{\gamma\delta}S_{\gamma\delta 1\dots 1}\,.
\eea
The counterparts of $T_{\alpha\beta 1\dots 1}$ will play the central role in our construction. Let us summarize the procedure for building these objects:
\begin{enumerate}[(i)]
\item Apply the symmetrizations corresponding to a given Young tableau to the basis (\ref{CombinedTheta}). This produces a tensor $R_{a_1\dots a_L}$.
\item Construct the irreducible $SO(6)$--tensor $S_{a_1\dots a_L}$ corresponding a given representation by subtracting traces:
\bea\label{RSTone}
S_{a_1\dots a_L}=R_{a_1\dots a_L}-
\sum_{i,j} \delta_{a_ia_j}R^{(a_ia_j)}_{a_1\dots a_{i-1}a_{i+1}\dots a_{j-1}a_{j+1}\dots a_L}
\eea
The objects $R^{(a_ia_j)}_{a_1\dots a_{i-1}a_{i+1}\dots a_{j-1}a_{j+1}\dots a_L}$ are chosen to ensure that 
\bea\label{RSTtwo}
\sum_{b,c}\delta^{bc}S_{a_1\dots a_{i-1}ba_{i+1}\dots a_{j-1}ca_{j+1}\dots a_L}=0
\eea
for all pairs of $(i,j)$. 
\item Decompose $S_{a_1\dots a_L}$ into irreducible representations $T_{a_1\dots a_L}$ of $SO(5)$ using modifications of (\ref{RSTone})--(\ref{RSTtwo}):
\bea\label{RSTthree}
&&T_{a_1\dots a_L}=S_{a_1\dots a_L}-
\sum_{i,j} {\hat\delta}_{a_ia_j}S^{(a_ia_j)}_{a_1\dots a_{i-1}a_{i+1}\dots a_{j-1}a_{j+1}\dots a_L},\nn
&&\sum_{b,c}{\hat\delta}^{bc}T_{a_1\dots a_{i-1}ba_{i+1}\dots a_{j-1}ca_{j+1}\dots a_L}=0,\quad
{\hat\delta}_{ab}={\hat\delta}^{ab}={\delta}_{ab}-\delta_{a1}\delta_{b1}
\eea
\item A Young tableau with $L_3\ne 0$ gives rise to two inequivalent irreducible tensors, $T$ described above and $(\star T)$ obtained using the definition (\ref{DefDualT}). 
\end{enumerate}
Let us now present some details of this construction along with several examples. 

\bigskip

We begin with recalling the standard correspondence between the Young tableaux and symmetrizations that allows one to construct tensors $R_{a_1\dots a_L}$ from the basis (\ref{CombinedTheta}). Starting with 
a Young tableau with marked boxes, one first antisymmetrizes the elements of every column, and then applies symmetrization within every row. For example\footnote{Following notation of \cite{Ma}, we denote an operator corresponding to a given Young tableau by ${\hat{\cal Y}}$.}, 
\bea\label{TExamples}
&&\byt a\\ b\eyt:\  R^{[110]}_{ab}={\hat{\cal Y}}\theta_{ab}=\frac{1}{\sqrt{2}}\left[\theta_{ab}-\theta_{ba}\right],\quad 
\byt a&b\eyt:\ 
R^{[200]}_{ab}={\hat{\cal Y}}\theta_{ab}=\frac{1}{\sqrt{2}}\left[\theta_{ab}+\theta_{ba}\right],\nn
&&
\byt a&c\\ b\eyt:\ 
R^{[210]}_{abc}={\hat{\cal Y}}\theta_{abc}=\frac{1}{2}\left\{[\theta_{abc}-\theta_{bac}]+[\theta_{cba}-\theta_{bca}]\right\},
\\
&&\byt a&b&c\eyt:\ 
R^{[300]}_{abc}={\hat{\cal Y}}\theta_{abc}=\frac{1}{\sqrt{6}}\left\{\theta_{abc}+perm\right\}\,.\nonumber
\eea
Note that a given diagram may correspond to several equivalent representations which are related by permutation of indices $\{i_1\dots i_L\}$ in (\ref{CombinedTheta}). For example, the diagram in the second row of (\ref{TExamples}) gives rise to two equivalent but distinct representations:
\bea\label{TExamplesAmbig}
\byt a&c\\ b\eyt:\ 
R^{[210]}_{abc}={\hat{\cal Y}}\theta_{abc},\quad
{\hat{\cal Y}}\theta_{acb}=\frac{1}{2}\left\{[\theta_{acb}-\theta_{bca}]+[\theta_{cab}-\theta_{bac}]\right\}. 
\eea
Clearly the second tensor is obtained from the first one by the interchange of $i_2$ and $i_3$ (see (\ref{CombinedTheta})). The main goal of this section is the group--theoretic construction of the wavefunction for $SO(6)/SO(5)$ coset, and as we will see below, only one representation with a given Young tableau (e.g., only $R^{[210]}_{abc}$ in the last example) is needed for that. Thus we will focus on the prescription (\ref{TExamples}) and will not discuss the ambiguity (\ref{TExamplesAmbig}) in detail.

The irreducible $SO(6)$--tensors are obtained by imposing the tracelessness condition (\ref{RSTtwo}). For the examples (\ref{TExamples})--(\ref{TExamplesAmbig}) the results are\footnote{While most of this appendix is dedicated to $SO(6)$, expressions (\ref{SExamples}) are written for $SO(2n)$. 
The answers for $n\ne 3$ will be used in the Appendix \ref{SecAppExmpl}.}
\bea\label{SExamples}
&&S^{[110]}_{ab}=R^{[110]}_{ab},\quad 
S^{[200]}_{ab}=R^{[200]}_{ab}-\frac{1}{2n}\delta_{ab}R^{[200]}_{cc},\nn
&&S^{[300]}_{abc}=R^{[300]}_{abc}
-\delta_{ab}V^{[300]}_{c}-\delta_{ac}V^{[300]}_{b}-\delta_{bc}V^{[300]}_{a},\quad
V^{[300]}_{a}=\frac{1}{2n+2}R^{[300]}_{aee},\\
&&S^{[210]}_{abc}=R^{[210]}_{abc}
-\delta_{ab}V^{[210]}_{c}-\delta_{bc}V^{[210]}_{a}-\delta_{ac}W^{[210]}_{b},\nn
&&\qquad V^{[210]}_{a}=\frac{2n R^{[210]}_{eea}-R^{[210]}_{eae}}{2n(2n+1)-2},\quad 
W^{[210]}_{a}=\frac{-2R^{[210]}_{eea}+(2n+1)R^{[210]}_{eae}}{2n(2n+1)-2}\,.\nonumber
\eea

The irreducible $SO(5)$ tensors are labeled by Young tableaux with specified locations of index one. Such tables are parameterized by five integers:
\bea\label{LvaluesQQ}
(L_1,L_2,|L_3|,l_1,l_2).
\eea
The first three entries give the lengths of the rows, while $l_1$ and $l_2$ specify the numbers of empty boxes in the first and the second rows. Some examples are given in equation (\ref{ExamplesMarked}). Every irreducible representation $(L_1,L_2,|L_3|)$ of $SO(6)$ can be split into several blocks with fixed numbers (\ref{LvaluesQQ}) by specifying the locations of index one and imposing the tracelessness condition (\ref{RSTthree}). For example, $S^{[200]}$ from (\ref{SExamples}) gives rise to three sets of $T^{[L_1L_2L_3]}_{[l_1l_2]}$:
\bea
[T^{[200]}_{[00]}]_{11}=S^{[200]}_{11},\quad 
[T^{[200]}_{[10]}]_{\alpha 1}=S^{[200]}_{\alpha 1},\quad
[T^{[200]}_{[20]}]_{\alpha \beta}=
S^{[200]}_{\alpha\beta}-\frac{1}{5}\delta_{\alpha\beta}\delta^{\gamma\delta}S^{[200]}_{\gamma\delta}
\eea
Under the $SO(5)$ rotations, these objects transform as a scalar, a vector, and an irreducible rank-two tensor. Note that $R^{[200]}_{11}$ contributes to $S^{[200]}_{11}$, but cancels from 
$[T^{[200]}_{[20]}]_{\alpha \beta}$\,:
\bea
[T^{[200]}_{[20]}]_{\alpha \beta}=
R^{[200]}_{\alpha\beta}
-\frac{1}{5}\delta_{\alpha\beta}\delta^{\gamma\delta}R^{[200]}_{\gamma\delta}
\eea
This is an example of a general feature: expression for tensor $T$ with $k$ $SO(5)$ indices contains only components of $R$ with $k$ $SO(5)$ indices. 

Let us now discuss the duality operation. For every tensor $R$ we can define\footnote{For simplicity we assume that indices $(b_1,\dots b_{6-m})$ appear in the same column in 
$R_{b_1\dots b_{6-m} a_{m+1}\dots a_L}$. In general, to get a nontrivial $(\star R)$, one has to add a sum over all possible positions of 
$(b_1,\dots b_{6-m})$ in $R$, but the resulting expression is not illuminating.}
\bea\label{DefDualT}
(\star R)_{a_1\dots a_m a_{m+1}\dots a_L}=
\frac{i}{(6-m)!}\sum_b \eps_{a_1\dots a_m b_1\dots b_{6-m}}R_{b_1\dots b_{6-m} a_{m+1}\dots a_L}
\eea
The right--hand side vanishes unless a Young tableau for $R$ has at least $(6-m)$ rows. This duality allows one to focus only on the Young tableaux with less than four rows, and every diagram with three rows has a dual contribution which is also covered by (\ref{TExamples}). For example,
\bea\label{ExampleDual}
R^{[111]}_{123}=\frac{1}{\sqrt{6}}\{\theta_{123}-perm\}\quad \Rightarrow\quad
(\star R^{[111]})_{123}=\frac{i}{\sqrt{6}}\{\theta_{456}-perm\}=iR^{[111]}_{456}\,.
\eea
Since the Levi--Civita symbol is invariant under $SO(6)$, the duality operation (\ref{DefDualT}) commutes with rotations. This implies that the $SO(6)$ tensors can be decomposed into self--dual and anti--self--dual ones, and such objects never mix. Similar decomposition works for tensors $S$ and $T$ defined by (\ref{RSTtwo}) and (\ref{RSTthree}).

\bigskip

After this brief review of the correspondence between the irreducible tensors and Young tableaux, we now turn to the wavefunctions for the $SO(6)/SO(5)$ coset. The relevant eigenvalues are given by (\ref{UndfEigValS5}), and to build the eigenfunctions, we need to evaluate the matrix elements of the product $g^L$, with $g$ given by (\ref{CosetG1}), in the basis of irreducible tensors. Such elements can be found by writing the irreducible tensors 
$T_{a_1\dots a_L}$ as linear combinations of $\theta_{a_1\dots a_L}$ and using the rule
\bea
\langle \theta_{a_1\dots a_L}|g^L|\theta_{b_1\dots b_L}\rangle\equiv
g_{a_1b_1}\dots g_{a_L b_L}\,.
\eea
Furthermore, we define the ``trace'' terms by the relation
\bea\label{SeparTrace}
\langle T_{a_1\dots a_L}|g^L|{T}'_{b_1\dots b_L}\rangle_{trace}\equiv 
\langle T_{a_1\dots a_L}|g^L|{T}'_{b_1\dots b_L}\rangle-
\langle R_{a_1\dots a_L}|g^L|{R}'_{b_1\dots b_L}\rangle\,,
\eea
where $(R,{R'})$ are the standard tensors associated with a given Young tableaux, and $(T,{T'})$ are traceless versions of the same tensors constructed using (\ref{RSTthree}). The matrix elements 
\bea\label{temp12}
\langle R_{a_1\dots a_L}|g^L|{R}'_{b_1\dots b_L}\rangle,\quad 
\langle T_{a_1\dots a_L}|g^L|{T}'_{b_1\dots b_L}\rangle
\eea
are equal to zero unless $R$ and ${R}'$ correspond to Young tableaux with the same shape. We will now show that the eigenfunctions of the Helmholtz equation (\ref{ScalarEqn}) can be obtained by contracting indices in the matrix elements (\ref{temp12}).

Starting with a particular Young tableau, we construct a set of diagonal matrix elements
\bea\label{MtrElenBefSum}
\langle T_{a_1\dots a_L}|g^L|{T}_{a_1\dots a_L}\rangle
\eea
These objects with different values of $(a_1\dots a_L)$ transform into each other under $SO(6)$ rotations, in particular, the individual matrix element (\ref{MtrElenBefSum}) is not invariant under $SO(5)$ unless all indices are equal to one. Since in the gauged WZW model, $SO(5)$ is a redundancy, only singlets under this group are allowed in the spectrum, so all indices not equal to one must be summed over. In other words, the wavefunction must have the form 
\bea\label{hatPsi}
{\hat\Psi}=\sum_{\alpha_k=2\dots 6} \langle 
T_{\alpha_1\dots \alpha_{n_1}1\alpha_{n_1+2}\dots \alpha_{n_2}1\alpha_{n_1+2}\dots}|g^L|
T_{\alpha_1\dots \alpha_{n_1}1\alpha_{n_1+2}\dots \alpha_{n_2}1\alpha_{n_1+2}\dots}\rangle
\eea
The insertions of index one are controlled by the ``marked'' Young tableau. By separating the trace contributions introduced in (\ref{SeparTrace}), we define
\bea\label{hatPhi}
{\hat\Phi}=\sum_{\alpha_k=2\dots 6} \langle 
R_{\alpha_1\dots \alpha_{n_1}1\alpha_{n_1+2}\dots \alpha_{n_2}1\alpha_{n_1+2}\dots}|g^L|
R_{\alpha_1\dots \alpha_{n_1}1\alpha_{n_1+2}\dots \alpha_{n_2}1\alpha_{n_1+2}\dots}\rangle\,.
\eea
Note that due to the summation of indices, the equivalent but distinct representations, such as ones given by the example (\ref{TExamplesAmbig}), lead to the same wavefunction. This justifies our focus on a single $R$ and $T$ for every class of equivalent representations.  Equations (\ref{hatPsi}) and (\ref{hatPhi}), along with definitions of $S$ and $T$ introduced earlier in this section, reproduce the expressions for ${\hat\Psi}$ and ${\hat\Phi}$ from  (\ref{TensorPrescribe}). 

So far we have focused on wavefunctions constructed from the matrix elements (\ref{MtrElenBefSum}), but for Young tableaux with nontrivial third row, one can define the dual irreducible tensors by (\ref{DefDualT}) with $m=3$ and consider additional matrix elements
\bea\label{MtrElemDual}
\langle (\star T)_{a_1\dots a_L}|g^L|T_{a_1\dots a_L}\rangle,\quad
\langle (T)_{a_1\dots a_L}|g^L|{\star T}_{a_1\dots a_L}\rangle,\quad
\langle (\star T)_{a_1\dots a_L}|g^L|(\star{T})_{a_1\dots a_L}\rangle
\eea
The $SO(5)$--invariant states can be constructed by contacting indices in each of these expressions, as in (\ref{hatPsi}). Since the dual states $(\star T)$ are obtained by re-arranging the components of the original states\footnote{This was illustrated by the example (\ref{ExampleDual}), but the statement is true in general.} $T$, contraction of the last expression in (\ref{MtrElemDual}) leads to a linear combination of wavefunctions (\ref{hatPsi}). Furthermore, contractions of the first two expressions in (\ref{MtrElemDual}) lead to the same set of wavefunctions, so without loss of generality, we can focus on contributions from the first term:
\bea\label{hatPsiDual}
&&{\tilde\Psi}=\sum_{\alpha_k=2\dots 6} \langle 
(\star T)_{\alpha_1\dots \alpha_{n_1}1\alpha_{n_1+2}\dots \alpha_{n_2}1\alpha_{n_1+2}\dots}|g^L|
T_{\alpha_1\dots \alpha_{n_1}1\alpha_{n_1+2}\dots \alpha_{n_2}1\alpha_{n_1+2}\dots}\rangle\nn
&&{\tilde\Phi}=\sum_{\alpha_k=2\dots 6} \langle 
(\star R)_{\alpha_1\dots \alpha_{n_1}1\alpha_{n_1+2}\dots \alpha_{n_2}1\alpha_{n_1+2}\dots}|g^L|
R_{\alpha_1\dots \alpha_{n_1}1\alpha_{n_1+2}\dots \alpha_{n_2}1\alpha_{n_1+2}\dots}\rangle
\eea
This is the second and the last set of eigenfunctions appearing in (\ref{TensorPrescribe}). Thus we have demonstrated that equation (\ref{TensorPrescribe}) exhausts all $SO(5)$--invariant objects constructed from tensors of $SO(6)$. In addition to such tensors, one can define ``spinor wavefunctions'', which are discussed in the end of section \ref{SecUndeform}. The remaining part of this appendix is dedicated to the study of the wavefunctions (\ref{hatPsi}) and (\ref{hatPsiDual}). 

\subsection{Power counting from Young tableaux}
\label{SecAppCount}

In this subsection we prove that function ${\hat\Psi}$ has  properties 1-7 listed on page \pageref{TensorProperties}. Function ${\tilde\Psi}$ and the self--dual object $({\hat\Psi}+{\tilde\Psi})$ will be studied in the next subsection. We begin with analyzing the expression (\ref{TensorPrescribe}) for the character ${\hat\Phi}$,
\bea\label{TensorPrescribe1}
{\hat\Phi}^{[L_1,L_2,L_3]}_{[l_1,l_2]}=\sum_{i_1\dots i_L}\sum_P (-1)^{\sigma(P)} 
g_{i_1 i_{P[1]}}\dots g_{i_L i_{P[L]}},\quad
L=L_1+L_2+L_3\,,
\eea
and demonstrating that this quantity has the same seven properties listed on page \pageref{TensorProperties} as ${\hat\Psi}$. The contribution of traces to ${\hat\Psi}$ will be discussed in the end of this subsection. 

\bigskip

To verify the properties of the eigenfunction ${\hat\Phi}$, we rewrite the expression (\ref{CosetG1}) for $g$ in terms of coordinate $Y$:
\bea\label{CosetG1App}
g=\begin{bmatrix}
	1&0\\
	0&q_{ab}
\end{bmatrix} \begin{bmatrix}
	\frac{2}{Y^2}-1&\frac{2X_b}{Y^2}\\
	-\frac{2X_a}{Y^2}& \delta_{ab}-\frac{2X_a X_b}{Y^2}
\end{bmatrix}, \quad q_{ab}=[(1+A)(1-A)^{-1}]_{ab}.
\eea
Here we defined\footnote{The branch with $X_3=-\sqrt{Y^2-1-X_1^2-X_2^2}$ can be analyzed in the same fashion.}
\bea
X_3=\sqrt{Y^2-1-X_1^2-X_2^2}\,.
\eea
It is clear that the scaling of (\ref{TensorPrescribe1}) with $(\alpha,\beta,Y)$ does not depend on the values of $(X_1,X_2)$ so to make the proof of properties 1-3 more transparent, we set  $X_1=X_2=0$.\footnote{This technical assumption is made just for presentational purposes.} Then matrix $g$ becomes  block--diagonal,
\bea
&&g=\left[\begin{array}{c|c}{\hat g}&0\\
\hline
0&g_{\dot a\dot b}\end{array}\right],\quad 
{\hat g}=\left[\begin{array}{cc}
\frac{2}{Y^2}-1&\frac{2\sqrt{Y^2-1}}{Y^2}\\
-\frac{2\sqrt{Y^2-1}}{Y^2}&\frac{2}{Y^2}-1
\end{array}\right],
\eea
where dotted indices take values $(3,4,5,6)$. The eigenvalues of the symmetric $4\times 4$ block 
$g_{\dot a\dot b}$ are given by
\bea
&&\mbox{eig}(g_{\dot a\dot b})=\Big\{\frac{i+a}{i-a},\frac{i-a}{i+a},
\frac{i+b}{i-b},\frac{i-b}{i+b}\Big\}.\nonumber
\eea
Every term in the right--hand side of (\ref{TensorPrescribe1}) has the structure
\bea\label{temp31}
[g_{11}]^t \Big[\prod (\text{Tr}[h^k])^{s_k}\Big] [\prod (g_{1a}[h^k]_{ab} g_{b1})^{r_k}\Big],
\eea
where
\bea
h_{ab}=\left[\begin{array}{c|c}\frac{2}{Y^2}-1&0\\
\hline
0&g_{\dot a\dot b}\end{array}\right],\quad g_{1a}=\frac{2\sqrt{Y^2-1}}{Y^2}[1,0,0,0,0]\,.
\nonumber
\eea
In particular, this implies that the expression (\ref{temp31}) is a polynomial in $Y^{-2}$ since $g_{12}$ is always multiplied by $g_{21}$. Furthermore, since $\mbox{Tr}[(g_{\dot a\dot b})^n]$ is a polynomial in variables $(\alpha,\beta)$ defined by (\ref{DefineABY}), so is (\ref{temp31}). This proves property 2 on page \pageref{TensorProperties}.  Let us now determine the degrees of the polynomials in the right hand side of (\ref{TensorPrescribe1}). 

The power of $Y$ comes from the $(1,2)$ block of the matrix $g_{ij}$:
\bea
{\hat g}=\left[\begin{array}{cc}
\frac{2}{Y^2}-1&\frac{2\sqrt{Y^2-1}}{Y^2}\\
-\frac{2\sqrt{Y^2-1}}{Y^2}&\frac{2}{Y^2}-1
\end{array}\right],\quad {\hat g}{\hat g}^T=I,\quad \mbox{det}(\hat g)=1.
\eea
The determinant condition implies that the leading power $\frac{1}{Y^{2q}}$ survives symmetrization in 
$[{\hat g}_{11}]^s[{\hat g}_{22}]^{q-s}$ if and only if all ${\hat g}_{11}$ and ${\hat g}_{22}$ appear in the same row. Thus we conclude that the leading power $p$ must be equal to the length of the longest row of the Young tableaux:
\bea\label{PropOne}
p= L_1.
\eea
This proves property 1 on page \pageref{TensorProperties}.
Furthermore, for large $(\alpha,\beta)$, the expression $\mbox{Tr}[({g}_{\dot a\dot b})^n]$ scales as 
$\alpha^n$ or 
$\beta^n$, so the character of the representation corresponding to a Young tableau with $L$ boxes scales as
\bea
\frac{1}{Y^{2p}}\alpha^{s_1}\beta^{s_2},\quad p+s_1+s_2=L. 
\eea
Combining this with property (\ref{PropOne}), we conclude that 
\bea
m_p=L_2+L_3.
\eea
The highest power of $\alpha$ for a given $p$ comes from the terms proportional to 
$[{\tilde g}_{22}]^p[{\tilde g}_{33}]^s$ (here ${\tilde g}$ is the diagonalized form of the symmetric matrix $g_{\dot a\dot b}$),  and such expressions survive symmetrization if and only if $s\le L_2$. This proves property 3 on page \pageref{TensorProperties}. 

To count the powers of $(x,y)$ we restore $(X_1,X_2)$ in (\ref{CosetG1}), but set $a=b=1$. This makes the discussion more transparent without modifying the counting. The expression (\ref{CosetG1}) becomes
\bea\label{temp26}
g=J+\frac{2}{Y^2}V W^T,\quad V=\{1,-X_3,-X_1,0,-X_2\},\quad W=\{1,X_3,X_1,0,X_2\}
\eea
Here $J$ is a constant matrix, which does not contribute to the wavefunction ${\hat\Phi}$ in the leading order in $Y^{-2}$. This leads to drastic simplifications in (\ref{temp31}), which now reduces to a product of scalars $\frac{2}{Y^2}(W^T V)$ and $\frac{2}{Y^2}$. In particular, note that while $X_3$ contains a square root, $(W^T V)$ is a polynomial in $(X_1,X_2)$, so ${\hat\Phi}$ is a polynomial in $(x,y)$. 

Substituting (\ref{temp26}) into (\ref{TensorPrescribe}), we conclude that any antisymmetrization eliminates the $Y$--dependent contribution since 
\bea
g_{mm}g_{nn}-g_{mn}g_{nm}=J_{mm}J_{nn}-J_{mn}J_{nm}\quad \mbox{(no summation)}
\eea
Thus the power of $Y^{-2}$ is equal to the maximal number of symmetrizations, and we recover (\ref{PropOne}). The power counting for $(X_1,X_2)$ works in the same way, but only indices not equal to one contribute\footnote{Recall the scaling $V_1=W_1=1$, $V_a\sim z$, $W_a\sim z$ with large
$z\sim (Y_1,Y_2)$.}, so the maximal power of $(x,y)$ is
\bea
n_p=l_1.
\eea
This proves the property 4 for the function ${\hat\Phi}$, and property 5 can be demonstrated in the same fashion, after one restores the $(a,b)$--dependence.

\bigskip

To summarize, we have shown that function ${\hat\Phi}$ satisfies the scalings listed on page \pageref{TensorProperties}. As we discussed in the last subsection, the correct wavefunction is ${\hat\Psi}$ rather than ${\hat\Phi}$, and these two expressions are related by subtraction of various traces. We will now demonstrate that such subtraction does not affect the power counting, so the function ${\hat\Psi}$ satisfies the properties  listed on page \pageref{TensorProperties}.

As discussed in the section \ref{SecAppYoung}, functions ${\hat\Phi}$ and  ${\hat\Psi}$ are obtained by taking matrix elements (\ref{hatPsi}) and  (\ref{hatPhi}), where tensors $T$ and $R$ are related by subtraction of traces:
\bea
T_{i_1\dots i_n}=R_{i_1\dots i_n}-
\sum_{k,l}\delta_{i_ki_l}V^{(i_k,i_l)}_{i_1\dots i_{k-1}i_{k+1}\dots i_{l-1}i_{l+1}i_n}-
\sum_{k,l}{\hat\delta}_{i_ki_l}W^{(i_k,i_l)}_{i_1\dots i_{k-1}i_{k+1}\dots i_{l-1}i_{l+1}i_n}
\eea
The tensors $(V,W)$, whose explicit form is not important for our discussion, are chosen in such a way that 
$T_{i_1\dots i_n}$ is traceless over any pair of indices both in $SO(6)$ and in $SO(5)$:
\bea\label{tmpTRSLSS}
\sum_{j=1}^6 T_{i_1\dots i_{p-1}j i_{p+1}\dots i_qj i_{q+1}\dots i_n}=0,\quad
\sum_{a=2}^6 T_{i_1\dots i_{p-1}a i_{p+1}\dots i_qa i_{q+1}\dots i_n}=0.
\eea
Schematically, the expression (\ref{hatPsi}) for ${\hat\Psi}$ can be written as
\bea
{\hat\Psi}&=&\langle T|g^L|T\rangle=\langle R-\delta V-{\hat\delta}W|g^L|R-\delta V-{\hat\delta}W\rangle\nn
&=&{\hat\Phi}-\langle T|g^L|\delta V+{\hat\delta}W\rangle-\langle \delta V+{\hat\delta}W|g^L|R\rangle\,.
\eea
Here we suppressed all indices present in  (\ref{hatPsi}) and (\ref{hatPhi}) to focus on the structure of the formula. According to the condition (\ref{tmpTRSLSS}), $T$ is traceless under contraction with both $\delta$ and ${\hat\delta}$. Then the summation implied in the last expression leads to an equality
\bea
\langle T|g^L|\delta V+{\hat\delta}W\rangle=0.
\eea 
Thus the relation between ${\hat\Psi}$ and ${\hat\Phi}$ can be written as
\bea\label{tmpPhiPsi}
{\hat\Psi}={\hat\Phi}-\langle \delta V|g^L|R\rangle-\langle {\hat\delta}W|g^L|R\rangle\,.
\eea
The contraction induced by $V$ is taken in the group, and it leads to a term
\bea 
\sum_{i=1}^{6}g_{im}g_{in}=\delta_{mn}.
\eea 
In power counting for ${\hat\Phi}$, every instance of $g$ contributed $Y^{-2}$. Then the last equation implies that the second term in the right--hand side of (\ref{tmpPhiPsi}) is subleading in the small--$Y$ limit, so it does not affect the power counting.

The third term in the right--hand side of (\ref{tmpPhiPsi}) involves a trace over the subgroup (recall that ${\hat\delta}$ is a projector that insures this), and its contribution is 
\bea 
\sum_{i=2}^{6}g_{im}g_{in}=\delta_{mn}-g_{1m}g_{1n}.
\eea 
According to the general expression (A.26), this contraction increases $r_k$ but decreases $s_k$, so it is subleading in 
$(\alpha,\beta)$. Thus the relation (\ref{tmpPhiPsi}) ensures that the power counting for function ${\hat\Phi}$ derived earlier in this subsection extends to function ${\hat\Psi}$ as well.

\bigskip

To summarize, in this subsection we have demonstrated that the eigenfunction ${\hat\Psi}$ defined by (\ref{TensorPrescribe}), (\ref{hatPsi}) satisfies all properties listed on page \pageref{TensorProperties}. In the next subsection similar features will be demonstrated for ${\tilde\Psi}$, completing the proof of the power counting summarized on page \pageref{TensorProperties}.

\subsection{Self--dual representations}
\label{SecAppDual}

Properties 1-7 listed on page \pageref{TensorProperties} for function 
${\tilde\Psi}$ can be proven using the arguments similar to the ones presented in subsection \ref{SecAppCount}. Rather than following this route, in this subsection focus on power counting in ${\hat\Psi}\pm{\tilde\Psi}$, then the properties of ${\tilde\Psi}$ will follow by combining the discussion of this and the last subsections. 

We begin with considering two representations with $L_3=0$. Since ${\tilde\Psi}$ vanishes for $L_3=0$, the corresponding wavefunctions are given by 
\bea\label{tempJul5}
{\hat\Psi}^{[L'_1,L'_2,0]}_{[l'_1,l'_2]}\quad\mbox{and}\quad {\hat\Psi}^{[L''_1,L''_2,0]}_{[l''_1,l''_2]}\,.
\eea
The product of these two functions correspond to a reducible representation, and it can be decomposed into irreducible pieces using the standard fusion rules. This provides an alternative way of proving properties 1-7 for ${\hat\Psi}$ with $L_3=0$ using induction:
\begin{enumerate}[(i)]
\item Perform a direct check of properties 1-7 for functions with 
\bea
(L_1,L_2,l_1,l_2)=(1,0,0,0),\ (1,0,1,0),\ (1,1,1,0),\ (1,1,1,1),\qquad L_3=0.\nonumber
\eea
\item Assume that such properties hold for all allowed values of $(L_1,L_2,l_1,l_2)$, such that 
\bea
L_1<N_1,\quad L_2<N_2,\quad  l_1<n_1,\quad l_2<n_2,\quad \mbox{where}\quad
N_1\ge n_1\ge N_2\ge n_2\ge 0.\nonumber
\eea
\item Consider a product of two functions (\ref{tempJul5}) with 
\bea
L_1'+L_1''=N_1,\quad L_2'+L_2''=N_2,\quad l_1'+l_1''=n_1,\quad l_2'+l_2''=n_2,\nonumber
\eea
and decompose it into irreducible representations. By the assumption (ii), the product has the structure (\ref{ConjPrprt1}) with 
\bea\label{tempJul5a}
p=L_1'+L_1''=N_1,\quad k=n_1+n_2,\quad m_p=N_2,\quad n_p=n_1.
\eea
According to (ii), all irreducible representations contributing to the product, with a possible exception of 
$(N_1,N_2,n_1,n_2)$ have the structure (\ref{ConjPrprt1}) with 
\bea
p<N_1,\quad k<n_1+n_2,\quad m_p<N_2,\quad n_p<n_1.\nonumber
\eea
This implies that the wavefunction ${\hat\Psi}^{[N_1,N_2,0]}_{[n_1,n_2]}$ has the structure (\ref{ConjPrprt1}) with powers (\ref{tempJul5a}), completing the proof by induction. 
\end{enumerate}
Extension of such induction to ${\hat\Psi}$ with non--zero values of $L_3$ is somewhat cumbersome, so we will not present it here since the result was proven in the last subsection using different logic.  

\bigskip

For non--zero values of $L_3$ we define
\bea
\Psi^{[L_1,L_2,M]}_{l_1,l_2}={\hat\Psi}^{[L_1,L_2,M]}_{l_1,l_2}+{\tilde \Psi}^{[L_1,L_2,M]}_{l_1,l_2},\quad
\Psi^{[L_1,L_2,-M]}_{l_1,l_2}={\hat\Psi}^{[L_1,L_2,M]}_{l_1,l_2}-{\tilde \Psi}^{[L_1,L_2,M]}_{l_1,l_2},
\eea
where $M=|L_3|$. Let us now use the induction to demonstrate that 
\bea\label{PsiL3App}
\Psi^{[L_1,L_2,L_3]}_{l_1,l_2}&=&\frac{1}{Y^{2p}}
[P^{(p)}_{m_p,n_p;k}+\gamma {\tilde P}^{(p)}_{m_p,{\tilde n}_p;k}]+
\frac{1}{Y^{2(p-1)}}[P^{(p-1)}_{m_{p-1},n_{p-1};k}+\gamma {\tilde P}^{(p-1)}_{m_{p-1},{\tilde n}_{p-1};k}]
+\dots\nn
&&\dots +
[P^{(0)}_{m_{0},n_{0};k}+\gamma {\tilde P}^{(0)}_{m_{0},{\tilde n}_{0};k}],
\eea 
where the powers of the polynomials obey (\ref{CounPowSmry}), and $\gamma$ is defined by 
\bea
\gamma=iab\sqrt{Y^2-1-X_1^2-X_2^2}
\eea
Furthermore, for large values of $(\alpha,\beta)\sim {\cal A}$ and fixed $(x,y)$, there is a scaling
\bea\label{ScalingApp}
\Big[P^{(p)}_{m_p,n_p;k}+\gamma{\tilde P}^{(p)}_{m_p,{\tilde n}_p;k}\Big]_{Y=0}\sim {\cal A}^{L_2+L_3}
\eea
Before starting the proof by induction, we observe that unlike other quantum number which have complicated fusion rules, $L_3$ just adds as a simple $U(1)$ charge when two wavefunctions are multiplied. We also present two explicit wavefunctions, which will be used as a starting point of the induction:
\bea 
&&\Psi^{[11\pm1]}_{[11]}=-5+6(\alpha+\beta)-8\alpha\beta+\frac{2}{Y^2}[3-4(\alpha+\beta)+(3-4\beta)x+(3-4\alpha)y]\nonumber \\
&&\quad\quad~~~~ +\frac{16\alpha\beta}{Y^2}(1\pm \gamma).
\eea 
Now the induction procedure follows the same steps as before:
\begin{enumerate}[(i)]
\item Perform a direct check of properties (\ref{PsiL3App}), (\ref{CounPowSmry}) (\ref{ScalingApp}) for functions with 
\bea
\left[\begin{array}{c}L_1,L_2,L_3\\
l_1,l_2\end{array}\right]=
\left[\begin{array}{c}1,0,0\\
0,0\end{array}\right], \left[\begin{array}{c}1,0,0\\
1,0\end{array}\right], \left[\begin{array}{c}1,1,0\\
1,0\end{array}\right],\left[\begin{array}{c}1,1,0\\
1,1\end{array}\right],\left[\begin{array}{c}1,1,\pm 1\\
1,1\end{array}\right].
\nonumber
\eea
\item Assume that such properties hold for all allowed values of $(L_1,L_2,L_3,l_1,l_2)$, such that 
\bea
L_1<N_1,\ L_2<N_2,\  |L_3|< N_3,\  l_1<n_1,\ l_2<n_2,\ \mbox{where}\
N_1\ge n_1\ge N_2\ge n_2\ge N_3.\nonumber
\eea
\item Consider a product of two functions
\bea\label{tempJul5Prod}
{\Psi}^{[L'_1,L'_2,L_3']}_{[l'_1,l'_2]}{\Psi}^{[L''_1,L''_2,L_3'']}_{[l''_1,l''_2]}
\eea
with 
\bea
L_1'+L_1''=N_1,\quad L_2'+L_2''=N_2,\quad L_3'+L_3''=N_3, \quad l_1'+l_1''=n_1,\quad l_2'+l_2''=n_2,\nonumber
\eea
and decompose it into irreducible representations. By the assumption (ii), the product (\ref{tempJul5Prod}) has the structure  (\ref{PsiL3App}) with 
\bea\label{tempJul5ab}
p=L_1'+L_1''=N_1,\quad k=n_1+n_2,\quad m_p=N_2,\quad n_p=n_1
\eea
and power $N_2+N_3$ in (\ref{ScalingApp}). 
According to (ii), all irreducible representations contributing to the product (\ref{tempJul5Prod}), with a possible exception of 
$(N_1,N_2,n_1,n_2)$ have the structure (\ref{PsiL3App}) with 
\bea
p<N_1,\quad k<n_1+n_2,\quad m_p<N_2,\quad n_p<n_1.\nonumber
\eea
This implies that the wavefunction ${\Psi}^{[N_1,N_2,N_3]}_{[n_1,n_2]}$ has the structure (\ref{PsiL3App}) with powers (\ref{CounPowSmry}), completing the inductive proof of relations (\ref{PsiL3App}), (\ref{CounPowSmry}).
\item
Since $L_3$ charges simply add, every irreducible term contributing to the product (\ref{tempJul5Prod}) has 
\bea
L_3=L_3'+L_3''.
\eea
This implies that all irreducible representations contributing to the product (\ref{tempJul5Prod}), with a possible exception of $(N_1,N_2,N_3,n_1,n_2)$, obey the scaling
\bea
\Big[P^{(p)}_{m_p,n_p;k}+\gamma{\tilde P}^{(p)}_{m_p,{\tilde n}_p;k}\Big]_{Y=0}\sim {\cal A}^{Q},\quad
Q=L_2+L_3<N_2+N_3\nonumber
\eea
Since the product itself has such scaling with $Q=N_2+N_3$, we conclude that the ingredients of the function ${\Psi}^{[N_1,N_2,N_3]}_{[n_1,n_2]}$ must obey
\bea
\Big[P^{(p)}_{m_p,n_p;k}+\gamma{\tilde P}^{(p)}_{m_p,{\tilde n}_p;k}\Big]_{Y=0}\sim {\cal A}^{N_2+N_3}.\nonumber
\eea
The last relation completes the inductive proof of the scaling (\ref{ScalingApp}). 
\end{enumerate}
After completing the proof of properties (\ref{PsiL3App}), (\ref{CounPowSmry}) (\ref{ScalingApp}) of the  function ${\Psi}^{[N_1,N_2,N_3]}_{[n_1,n_2]}$, let us now analyze the implications for 
${\hat\Psi}^{[L_1,L_2,M]}_{[l_1,l_2]}$ and ${\tilde\Psi}^{[L_1,L_2,M]}_{[l_1,l_2]}$. First we observe that in the physical domain of coordinates\footnote{Note that the scaling (\ref{ScalingApp}) explores an unphysical regime since large values of $(\alpha\beta)$ with fixed $(x,y)$ give imaginary 
$X_3=\sqrt{Y^2-1-X_1^2-X_2^2}\rightarrow \sqrt{Y^2-1}$ at $Y=0$. In this regime both
${\hat\Psi}^{[L_1,L_2,M]}_{[l_1,l_2]}$ and ${\tilde\Psi}^{[L_1,L_2,M]}_{[l_1,l_2]}$ are real, but the relations (\ref{PsiL3AppXX}) still hold due to their analyticity.} polynomials $P$ and ${\tilde P}$ in (\ref{PsiL3App}) are real and $\gamma$ is imaginary. Also, due to $i$ in the definition (\ref{DefDualT}) of the dual tensor real wavefunctions ${\hat\Psi}$ from (\ref{hatPsi}) give rise to imaginary ${\tilde\Psi}$ in (\ref{hatPsiDual}). Then taking a real and imaginary parts of $\Psi^{[L_1,L_2,M]}_{l_1,l_2}$, we find
\bea\label{PsiL3AppXX}
{\hat\Psi}^{[L_1,L_2,M]}_{l_1,l_2}&=&\frac{P^{(p)}_{m_p,n_p;k}}{Y^{2p}}
+
\frac{P^{(p-1)}_{m_{p-1},n_{p-1};k}}{Y^{2(p-1)}}
+\dots+
P^{(0)}_{m_{0},n_{0};k},\\
{\tilde\Psi}^{[L_1,L_2,L_3]}_{l_1,l_2}&=&\frac{\gamma}{Y^{2p}}
{\tilde P}^{(p)}_{m_p,{\tilde n}_p;k}+
\frac{\gamma}{Y^{2(p-1)}}{\tilde P}^{(p-1)}_{m_{p-1},{\tilde n}_{p-1};k}
+\dots+
\gamma {\tilde P}^{(0)}_{m_{0},{\tilde n}_{0};k}\,.\nonumber
\eea 
This justifies the expansions (\ref{ConjPrprt1}), (\ref{ConjPrprt1a}) and completes the proof of properties 1-7 listed on page \pageref{TensorProperties}.

\section{Examples of wavefunctions on deformed cosets}
\label{SecAppExmpl}

In section \ref{SecDeformed} and \ref{SecS3} we discussed the structure of eigenstates of the Helmholtz equation (\ref{ScalarEqn}) on deformed spheres, and in this appendix we present some explicit examples to demonstrate that they fit the general pattern. We begin with constructing the wavefunctions for the standards $SO(6)/SO(5)$ coset and showing that they satisfy the seven properties listed on page \pageref{TensorProperties}. Then in section \ref{SecAppEx2} we discuss the effect of the deformation and show that it fits into the general structure uncovered in section \ref{SecDeformed}. Some examples of wavefunctions on the deformed three--dimensional sphere are discussed in section \ref{SecAppEx3}.

\subsection{Eigenfunctions for the $SO(6)/SO(5)$ gauged WZW model}
\label{SecAppEx1}
\renewcommand{\theequation}{B.\arabic{equation}}
\setcounter{equation}{0}

Let us present some solutions of the Helmholtz equation (\ref{ScalarEqn}) on the $SO(6)/SO(5)$ coset and demonstrate that all examples fit into the general pattern discussed on page \pageref{TensorProperties}. As we demonstrated in section \ref{SecUndeform}, the eigenvalues of the Helmholtz equation are given by (\ref{UndfEigValS5}), and they are characterized by five (half--)integer numbers $(L_1,L_2,L_3,l_1,l_2)$ satisfying inequalities listed in (\ref{UndfEigValS5}). Most examples presented in this appendix correspond to small values of all five parameters.

Let us begin with an infinite set of examples with arbitrary $L_1$ and with 
\bea\label{L1Ne0}
L_2=L_3=l_1=l_2=0.
\eea
According to the discussion presented in section \ref{SecUndeform}, the corresponding wavefunction depends only on one argument $Y$, so we impose an ansatz:
\bea
\Psi^{[L_1,0,0]}_{[0,0]}=f(z),\qquad z\equiv \frac{1}{Y^{2}}.
\eea
Substitution into the the Helmholtz equation (\ref{ScalarEqn}) leads to an ODE:
\bea
4z(1-z)f''+10(1-2z)f'+\Lambda f=0,
\eea
and the solution regular at $z=0$ can be expressed in terms of the hypergeometric function:
\bea\label{FHyper100}
\Psi^{[L_1,0,0]}_{[0,0]}=F\left[-\nu,\nu+4;\frac{5}{2};z\right], \qquad \Lambda=4\nu(\nu+4).
\eea
Normalizable solutions are described by polynomials in $z$, then $\nu$ must be a non--negative integer, and the relation (\ref{UndfEigValS5}) is recovered upon identification $L_1=\nu$. Here are some explicit examples of the wavefunctions (\ref{FHyper100}):
\bea
\Psi^{[100]}_{[00]}=\frac{2}{Y^2}-1,\ 
\Psi^{[200]}_{[00]}=\frac{5}{24}-\frac{1}{Y^2}+\frac{1}{Y^4},\
\Psi^{[300]}_{[00]}=\frac{1}{Y^6}-\frac{3}{2 Y^4}+\frac{21}{32 Y^2}-\frac{5}{64}\,.
\eea
Note that the condition (\ref{L1Ne0}) restricts to tensor representations where all five parameters 
(\ref{Lvalues}) are integers. 

Continuing with tensor representations, we now look at small values of $[L_1,L_2,L_3]$. The $[1,0,0]$ block contains only two states:
\bea\label{tmpXX}
\begin{ytableau}1\end{ytableau}:&&\Psi^{[100]}_{[00]}=\frac{2}{Y^2}-1,\nn
\begin{ytableau}
~
\end{ytableau}:&&\Psi^{[100]}_{[10]}=\frac{2}{Y^2}(1+2x+2y)-5+4(\alpha+\beta).
\eea 
To verify that all seven properties listed on page \pageref{TensorProperties} are satisfied, we observe that the second state in (\ref{tmpXX}) has
\bea
P^{(p)}_{m_p,n_p,k}=2(1+2x+2y)\ \Rightarrow\ p=1,\ m_p=0,\ n_p=1,\ k=1,
\eea
in agreement with (\ref{CounPowSmry}). For the $[2,0,0]$ block we find:
\bea\label{tmpXX1}
\begin{ytableau}1&1\end{ytableau}:&&\hskip -0.4cm\Psi^{[200]}_{[00]}=\frac{1}{Y^4}-\frac{1}{Y^2}+\frac{5}{24},\\
\begin{ytableau}~&1\end{ytableau}:&&\hskip -0.4cm\Psi^{[200]}_{[10]}=-\frac{16}{Y^4}(1+2x+2y)+\frac{8}{Y^2}(4-2\alpha-2\beta+x+y)-10+8(\alpha+\beta)\nonumber, \\
\begin{ytableau}~&~\end{ytableau}:&&\hskip -0.4cm\Psi^{[200]}_{[20]}=
\frac{256}{5Y^4}[1+5(x+y)(x+y+1)]+
32[7(1-2\alpha-2\beta)+8(\alpha^2+\alpha\beta+\beta^2)]\nonumber\\
&&\qquad+\frac{64}{5Y^2}[5\alpha+5\beta-7+(8\alpha+4\beta-7)x+(4\alpha+8\beta-7)y].\nonumber
\eea  
Again, relations (\ref{CounPowSmry}) are satisfied for all three states. The remaining Young tableaux with two boxes give rise to the following states:
\bea\label{tmpXX2}
\begin{ytableau}~\\1\end{ytableau}:&&\hskip -0.4cm
\Psi^{[110]}_{[10]}=\frac{4}{Y^2}[2(\alpha+\beta-1)-(x+y)]+5-4(\alpha+\beta),\\
\begin{ytableau}~\\~\end{ytableau}:&&\hskip -0.4cm\Psi^{[110]}_{[11]}=
\frac{32}{Y^2}[2(1-\alpha-\beta)+(3-4\beta)x+(3-4\alpha)y]
-16(5-6\alpha-6\beta+8\alpha\beta)\nonumber.
\eea 
Note that all examples discussed so far had $L_3=0$, so the wavefunctions were polynomials in 
$(\alpha,\beta,x,y)$, and $\tilde\Psi$ defined by (\ref{TensorPrescribe}) vanished. The simplest Young tableau with nontrivial 
$\tilde\Psi$ contains three boxes arranged in a column, and as discussed in section \ref{SecGroup}, one such diagram gives rise to two states:
\bea\label{ExmpThrRows}
\begin{ytableau}~\\~\\1\end{ytableau}:&&\Psi^{[11\pm1]}_{[11]}=\frac{16\alpha\beta}{Y^2}(1\pm \gamma)+\frac{2}{Y^2}[3-4(\alpha+\beta)+(3-4\beta)x+(3-4\alpha)y]\nonumber \\
&&\quad\quad~~~~ -5+6(\alpha+\beta)-8\alpha\beta.
\eea 
Recall that variable $\gamma$ was defined by (\ref{ConjPrprt1a}):
\bea
\gamma=iabX_3=ia b\sqrt{Y^2-X_2^2-X_2^2-1}\,.\nonumber
\eea
As expected, power counting in (\ref{ExmpThrRows}) agrees with (\ref{CounPowSmry}). Furthermore, the structure
\bea
\Psi^{[L_1,L_2,\pm M]}_{[l_1,l_2]}=A\pm\gamma B\,.
\eea
follows from the properties 1-7 discussed on page \pageref{TensorProperties}.

All examples presented so far correspond to Young tableaux which are shaped either as rows or as columns, and we conclude this subsection by writing some wavefunctions coming from Young tableaux with mixed symmetry:
\bea\label{tmpXX3}
\begin{ytableau}~&1\\1\end{ytableau}:&&\hskip -0.5cm
\Psi^{[210]}_{[10]}=\frac{48}{Y^4}[2(-1+\alpha+\beta)-(x+y)]\nonumber\\
&&\quad+\frac{96}{5Y^2}(\frac{11}{2}-5\alpha-5\beta+x+y)+\frac{24}{5}(-5+4\alpha+4\beta),\nn
\begin{ytableau}~&1\\~\end{ytableau}:&&\hskip -0.5cm
\Psi^{[210]}_{[11]}=\frac{96}{Y^4}[2(-1+\alpha+\beta)+(-3+4\beta)x+(-3+4\alpha)y]\\
&&\quad +\frac{32}{Y^2}[9+8\alpha\beta-10(\alpha+\beta)+(3-4\beta)x+(3-4\alpha)y]
-16[5+8\alpha\beta-6(\alpha+\beta)],\nonumber
\eea
\bea
\begin{ytableau}~&~\\1\end{ytableau}:&&\hskip -0.5cm
\Psi^{[210]}_{[20]}=
\frac{24}{Y^4}[2(-1+\alpha+\beta)+(-7+8\alpha+4\beta)x+(-7+4\alpha+8\beta)y-4(x+y)^2]\nn
&&\quad +\frac{24}{Y^2}[7(1-2\alpha-2\beta)+8(\alpha^2+\alpha\beta+\beta^2)-(-7+8\alpha+4\beta)x-(-7+4\alpha+8\beta)y],\nn
&&\quad-12[7(1-2\alpha-2\beta)+8(\alpha^2+\alpha\beta+\beta^2)]\nonumber\\
\begin{ytableau}~&~\\~\end{ytableau}:&&\hskip -0.5cm
\Psi^{[210]}_{[21]}=-\frac{8\cdot 96}{Y^4}\Big[(-3+4\beta)x^2+2(-3+2\alpha+2\beta)xy+(-3+4\alpha)y^2\Big]\nonumber\\
&&\quad-\frac{96}{Y^4}\Big[27(x+y)+16[(\alpha+2\beta)x+(2\alpha+\beta)y]+6(-1+\alpha+\beta)\Big]
\nonumber \\
&&\quad-\frac{96}{Y^2}[(2\alpha+2\beta-3)(8\alpha+8\beta-7)]\nn
&&\quad-\frac{192}{Y^2}[(4\beta-3)(8\alpha+4\beta-7)x+(4\alpha-3)(4\alpha+8\beta-7)y]\nonumber\\
&&\quad -48[64\alpha\beta(\alpha+\beta)-16(3\alpha^2+10\alpha\beta+3\beta^2)+84(\alpha+\beta)-35]
\nonumber
\eea
In the next subsection we will show how the wavefunctions presented here should be combined to solve the Helmholtz equation (\ref{ScalarEqn}) on the deformed geometry. 

\subsection{Eigenfunctions for the deformed $SO(6)/SO(5)$ coset}
\label{SecAppEx2}

As we demonstrated in section \ref{SecDefLapl}, the solutions of the Helmholtz equation (\ref{ScalarEqn}) on the deformed 
$SO(6)/SO(5)$ coset must come from linear combinations of the undeformed eigenfunctions corresponding to Young tableaux with the same shape. In this subsection we present the examples of such mixtures coming from the four families (\ref{tmpXX}), (\ref{tmpXX2}), (\ref{ExmpThrRows}), (\ref{tmpXX3}) discussed earlier. After writing the exact results, we will focus on simplifications that arise in the rescaling limit introduced in section \ref{SecDefResc}. 

\bigskip

Each member of the families (\ref{tmpXX}), (\ref{tmpXX2}), (\ref{ExmpThrRows}), (\ref{tmpXX3}) solves the Helmholtz equation (\ref{ScalarEqn}) on the $SO(6)/SO(5)$ coset, but deformation mixes the members of a given family. Using relations (\ref{LapOpe}), we write the deformed equation as 
\bea\label{LapOpeApp} 
\left[\Delta_0-\kappa\Delta_1\right]\Psi+\Lambda\frac{1-\lambda^2}{1+\lambda^2}\Psi=0
\eea
It is the differential operator $\Delta_1$ that leads to the mixing. Writing the wavefunction $\Psi$ as a linear combination of the undeformed solutions with fixed $(L_1,L_2,L_3)$,
\bea\label{DefPsiInBas}
\Psi=\sum_{l_1,l_2} c_{l_1,l_2}\Psi^{[L_1,L_2,L_3]}_{[l_1,l_2]}\,,
\eea
we find that equation (\ref{LapOpeApp}) reduces to a system of algebraic relations for constants $c_{l_1,l_2}$. Note that summation in (\ref{DefPsiInBas}) extends only over the allowed values of $(l_1,l_2)$ which satisfy inequalities (\ref{UndfEigValS5}). Let us now present some explicit examples.

The family (\ref{tmpXX}) contains only two members, then equation (\ref{LapOpeApp}) in the relevant sector becomes
\bea\label{tmpXXdef}
\begin{ytableau}
~
\end{ytableau}: \quad \begin{bmatrix}
	20&-4\kappa \\
	-20\kappa &4
\end{bmatrix}\begin{bmatrix}
	c_{00}\\c_{10}
\end{bmatrix}=\Lambda \begin{bmatrix}
	c_{00}\\c_{10}
\end{bmatrix}
\eea
The functions $\Psi^{[100]}_{[00]}$ and $\Psi^{[100]}_{[10]}$ are given by (\ref{tmpXX}). In the absence of deformation (i.e., for $\kappa=0$), the last equation reproduces the expected result, $\Lambda=(20,4)$. For non--zero values of $\kappa$, one can find the two eigenvalues and two eigenfunctions by solving the system (\ref{tmpXXdef}). In particular, the eigenvalues satisfy a quadratic equation with constant coefficients. 

The families (\ref{tmpXX}) (\ref{tmpXX1}), (\ref{tmpXX2}), and (\ref{tmpXX3}) lead to the following systems of linear equations:
\bea\label{tmpXX2def} 
\begin{ytableau}
  ~&~
\end{ytableau}:&& \begin{bmatrix}
	48&\kappa &0 \\
	\frac{2304}{5}\kappa &32&-\kappa\\
	0&\frac{896}{5}\kappa &8
\end{bmatrix}\begin{bmatrix}
	c_{00}\\c_{10}\\c_{11}
\end{bmatrix}=\Lambda \begin{bmatrix}
	c_{00}\\c_{10}\\c_{11}
\end{bmatrix}\nn
\begin{ytableau}
~\\
~
\end{ytableau}: && \begin{bmatrix}
	16& \kappa \\
	128\kappa &8
\end{bmatrix}\begin{bmatrix}
	c_{10}\\c_{11}
\end{bmatrix}=\Lambda\begin{bmatrix}
	c_{10}\\c_{11}
\end{bmatrix}\\
\begin{ytableau}
~&~\\
~
\end{ytableau}: &&\begin{bmatrix}
	44&-9\kappa &-4\kappa&0 \\
	-50\kappa &36&0&\kappa\\
	-\frac{35}{2}\kappa&0 &20&\frac{3}{4}\kappa \\
	0&126\kappa &120\kappa &12
\end{bmatrix}\begin{bmatrix}
	c_{10}\\c_{11}\\c_{20}\\c_{21}
\end{bmatrix}=\Lambda \begin{bmatrix}
	c_{10}\\c_{11}\\c_{20}\\c_{21}
\end{bmatrix}\nonumber
\eea 
The values of $\Lambda$ are determined by solving the appropriate characteristic equations, and as a consistency check, we observe 
that the expressions (\ref{UndfEigValS5}) are recovered for $\kappa=0$. This can be seen by focusing on the diagonal elements of three matrices in (\ref{tmpXX2def}). The wavefunctions 
$\Psi^{[1,1,\pm1]}$ are given by the undeformed expressions (\ref{ExmpThrRows}), and the corresponding eigenvalue is 
$\Lambda=12(1-\kappa)$, as expected from the general discussion surrounding equation (\ref{ProtectedStates}). 

Although the matrices with constant coefficients are relatively simple, the functional form of 
$\Psi^{[L_1,L_2,L_3]}_{[l_1,l_2]}$ becomes rather complicated as parameters become larger, as can be seen from  (\ref{tmpXX}), (\ref{tmpXX2}), (\ref{ExmpThrRows}), (\ref{tmpXX3}). Interestingly, the eigenvalues $\Lambda$ can be extracted without using the explicit form of $\Psi^{[L_1,L_2,L_3]}_{[l_1,l_2]}$. The relevant procedure was developed in section \ref{SecDefMatr} and summarized in steps (i)--(iii) on page \pageref{StepsS5}. Specifically the problem is reduced to finding eigenvalues of some matrix $M^{[L_1,L_2,L_3]}$, which is related to matrices appearing in (\ref{tmpXX2def}) by a change of basis. We refer to section \ref{SecDefMatr} for the details, and here we just present the results for the families (\ref{tmpXX}), (\ref{tmpXX2}), (\ref{ExmpThrRows}), (\ref{tmpXX3}):
\bea
\begin{ytableau}
~
\end{ytableau}:&&M^{[100]}=\begin{bmatrix}
	\Lambda-24&20(\kappa^2-1)\\
	4&\Lambda
\end{bmatrix}\nn
\begin{ytableau}
~&~
\end{ytableau}:&&M^{[200]}=
\begin{bmatrix}
\Lambda-64 & 28 \left(\kappa ^2-1\right) & 0 \\
 16 & \Lambda-24 & 48 \left(\kappa ^2-1\right) \\
 0 & 4 &\Lambda \\
\end{bmatrix}\nn
\begin{ytableau}
~\\
~
\end{ytableau}:&&
M^{[110]}=\begin{bmatrix}
 \Lambda+8 (\kappa -2) & 4 (\kappa +1) \\
 16 \kappa  & \Lambda-8 (\kappa +1) \\
\end{bmatrix}\\
\begin{ytableau}
~&~\\
~
\end{ytableau}:&&
M^{[210]}=\begin{bmatrix}
 \Lambda+16 (\kappa -3) & 4 (\kappa +1) & 28 \left(\kappa ^2-1\right) & 0 \\
 32 \kappa  & \Lambda-8 (2 \kappa +5) & 0 & 28 \left(\kappa ^2-1\right) \\
 4 & 0 & \Lambda+8 (\kappa -2) & 4 (\kappa +1) \\
 0 & 4 & 16 \kappa  & \Lambda-8 (\kappa +1) \\
\end{bmatrix}\nonumber
\eea 
The eigenvalues for every family are determined by solving the characteristic equation (\ref{detMS5}), which has degree $(L_1-L_2+1)(L_2-L_3+1)$ in $\Lambda$.

\subsection{Eigenfunctions for the deformed $SO(4)/SO(3)$ coset}
\label{SecAppEx3}

In this appendix we present the counterparts of the last two subsections for the deformed $SO(4)/SO(3)$ coset. Since the logic is identical to the one used before, we only quote the results. 

\bigskip

The eigenfunctions of the Helmholtz equation on the undeformed three--sphere are characterized by three (half-) integer quantum numbers $(L_1,L_2,l_1)$, and the eigenvalues are given by (\ref{UndfEigVal43}). The wavefunctions (\ref{UndfEigFnc43}) become rather complicated for large values of the quantum numbers, but one infinite family (the counterpart of (\ref{FHyper100})) can be easily constructed:
\bea\label{yeigens3}
\Psi^{[L_1,0]}_{[0]}=F\left[-L_1,L_1+2,\frac{3}{2};z\right],\quad z\equiv\frac{1}{Y^2}\quad \Lambda=4L_1(L_1+2)\,.
\eea 
Here are some explicit examples:
\bea 
\Psi^{[10]}_{[0]}=1-\frac{2}{Y^2},~\Psi^{[20]}_{[0]}=-\frac{8}{Y^4}+\frac{8}{Y^2}-\frac{3}{2},~\Psi^{[30]}_{[0]}=-\frac{32}{Y^6}+\frac{48}{Y^4}-\frac{20}{Y^2}+2.
\eea 

Next we discuss wavefunctions with small values of $(L_1,L_2,l_1)$ and show how they combine under the deformation via a counterpart of (\ref{DefPsiInBas}):
\bea
\Psi=\sum_{l_1} c_{l_1}\Psi^{[L_1,L_2]}_{[l_1]}\,,
\eea
We begin with the block $(L_1,L_2)=(1,0)$. The undeformed wavefunction are
\bea
&&\byt
1\eyt:\quad \Psi^{[10]}_{[0]}=\frac{2}{Y^2}-1,\nonumber\\
&&\byt
~\eyt:\quad \Psi^{[10]}_{[1]}=\frac{2}{Y^2}(2 x+1)-3+4 \alpha\,,
\eea 
and they clearly satisfy the properties 1-5 listed on page \pageref{PagePropS3}. The mixing comes from the counterpart of equation (\ref{tmpXXdef}):
\bea
\begin{ytableau}
~
\end{ytableau}: \quad \begin{bmatrix}
	12&-4\kappa \\
	-12\kappa &4
\end{bmatrix}\begin{bmatrix}
	c_{0}\\c_{1}
\end{bmatrix}=\Lambda \begin{bmatrix}
	c_{0}\\c_{1}
\end{bmatrix}
\eea
For the $[20]$ block we find:
\bea 
&&\byt 1&1\eyt: \quad \Psi^{[20]}_{[0]}=\frac{16}{3 Y^4}-\frac{16}{3 Y^2}+1,\\
&&\byt ~&1\eyt:\quad \Psi^{[20]}_{[1]}=\frac{8}{Y^4}(1+2x)-4\frac{3+x-2\alpha}{Y^2}+3-4\alpha,\nonumber\\
&&\byt ~&~\eyt:\quad \Psi^{[20]}_{[2]}=\frac{8}{3Y^4}(1+6x+6x^2)+
\frac{4}{Y^2}[2\alpha-\frac{5}{3}+(8\alpha-5)x]+5-20\alpha+16\alpha^2\,.\nonumber
\eea 
and the counterpart of the first equation in (\ref{tmpXX2def}) is
\bea 
\begin{bmatrix}
 32 & -\frac{32 \kappa }{3} & 0 \\
 -32 \kappa  & 24 & -8 \kappa  \\
 0 & -\frac{40 \kappa }{3} & 8 \\
\end{bmatrix}\begin{bmatrix}
	c_{0}\\c_{1}\\c_{2}
\end{bmatrix}=
\Lambda \begin{bmatrix}
	c_{0}\\c_{1}\\c_{2}
\end{bmatrix}
\eea 
There are two one--column representations:
\bea 
\byt ~\\1\eyt:\quad \Psi^{[1,\pm1]}_{[1]}=3-4\alpha-\frac{4}{Y^2}[1+x-2\alpha(1\pm\gamma)].
\eea 
and the wavefunctions do not mix after the deformation. Functions $\Psi^{[1,1,\pm1]}$ in the 
$SO(6)/SO(5)$ case had the same property. 

We conclude this appendix by giving explicit form of two matrices appearing in (\ref{detS3}): 
\bea 
M^{[10]}=\begin{bmatrix}
 \frac{\Lambda+4 \kappa -12}{4 \kappa +4} & -1 \\
 -\frac{\kappa }{2 (\kappa +1)} & \frac{\Lambda-4(\kappa +1)}{16 (\kappa +1)} \\
\end{bmatrix},\quad
M^{[20]}=\begin{bmatrix}
 \frac{\Lambda+16 \kappa -32}{4 \kappa +4} & -1 & 0 \\
 -\frac{2 \kappa }{\kappa +1} & \frac{\Lambda-8 (\kappa +3)}{16 (\kappa +1)} & -1 \\
 0 & -\frac{2 \kappa }{9 (\kappa +1)} & \frac{\Lambda-8(\kappa +1)}{36 (\kappa +1)} \\
\end{bmatrix}\,.
\eea

\end{document}